\documentclass[12pt]{article}  

\pdfoutput=1
                   
\usepackage[utf8]{inputenc}
\usepackage[a4paper,textheight=19.5cm,centering]{geometry}
\usepackage[sort,numbers]{natbib}
\usepackage{afterpage}

\usepackage{xcolor}
\usepackage{fix-cm,type1cm}
\usepackage{amsmath}
\usepackage{amsfonts}

\usepackage{graphicx}
\usepackage{stmaryrd}
\SetSymbolFont{stmry}{bold}{U}{stmry}{m}{n}

\usepackage[normalem]{ulem}

\newcommand{\be}{\begin{equation}}
\newcommand{\ee}{\end{equation}}
\newcommand{\vc}[1]{\mbox{\boldmath\ensuremath{#1}}}  
\newcommand{\ts}[1]{\mbox{\boldmath\ensuremath{#1}}}  
\newcommand{\ud}{\,\mathrm{d}} 
\newcommand{\jumpop}[1]{\llbracket #1 \rrbracket} 
\newcommand{\RE}{\mathrm{Re}} 
\newcommand{\DA}{\mathrm{Da}} 
\newcommand{\vf}{\phi} 
\newcommand{\svf}{\phi_s} 
\newcommand{\fvf}{{\phi_f}} 
\newcommand{\svfcrit}{{\phi_{sc}}} 
\newcommand{\stress}{\ts \tau} 
\newcommand{\stresscmpt}{\tau} 
\newcommand{\sstress}{{\ts{\tau}_s}} 
\newcommand{\fstress}{{\ts{\tau}_f}} 
\newcommand{\srt}{\ts{\dot \gamma}} 
\newcommand{\fsrt}{{\ts{\dot \gamma}_f}} 
\newcommand{\ssrt}{{\ts{\dot \gamma}_s}} 
\newcommand{\srtcmpt}{\dot \gamma} 

\newcommand{\vel}{\vc{u}} 
\newcommand{\svel}{{\vc{u}_s}} 
\newcommand{\fvel}{{\vc{u}_f}} 
\newcommand{\velu}{u} 
\newcommand{\velv}{v} 
\newcommand{\prs}{p} 
\newcommand{\fprs}{{p_f}} 
\newcommand{\cprs}{{p_c}} 

\newcommand{\velb}{{\vc{U}_B}} 
\newcommand{\velp}{{\tilde{\vel}}} 
\newcommand{\velup}{{\tilde{u}}} 
\newcommand{\velvp}{{\tilde{v}}} 
\newcommand{\vely}{{\hat{\vel}}} 
\newcommand{\veluy}{{\hat{u}}} 
\newcommand{\velvy}{{\hat{v}}} 
\newcommand{\prsp}{{\tilde{p}}} 

\newcommand{\fvfb}{{\Phi_f}} 
\newcommand{\fvfp}{{\tilde{\phi}_f}} 

\newcommand{\svfb}{{\Phi_s}} 
\newcommand{\svfp}{{\tilde{\phi}_s}} 

\newcommand{\fvelub}{{U_f}} 
\newcommand{\fvelup}{{\tilde{u}_f}} 
\newcommand{\fvelvp}{{\tilde{v}_f}} 

\newcommand{\svelub}{{U_s}} 
\newcommand{\svelup}{{\tilde{u}_s}} 
\newcommand{\svelvp}{{\tilde{v}_s}} 

\newcommand{\fprsb}{{P_f}} 
\newcommand{\fprsp}{{\tilde{p}_f}} 
\newcommand{\cprsb}{{P_c}} 
\newcommand{\cprsp}{{\tilde{p}_c}} 

\newcommand{\sstressb}{{T_s}} 
\newcommand{\sstressp}{{\tilde \tau_s}} 
\newcommand{\fstressb}{{T_f}} 
\newcommand{\fstressp}{{\tilde \tau_f}} 

\newcommand{\hp}{{\tilde h}} 

\newcommand{\svfy}{{\hat{\phi}_s}} 
\newcommand{\fveluy}{{\hat{u}_f}} 
\newcommand{\fvelvy}{{\hat{v}_f}} 

\newcommand{\jveluy}{{\hat{u}_j}} 
\newcommand{\jvelvy}{{\hat{v}_j}} 

\newcommand{\sveluy}{{\hat{u}_s}} 
\newcommand{\svelvy}{{\hat{v}_s}} 
\newcommand{\fprsy}{{\hat{p}_f}} 
\newcommand{\cprsy}{{\hat{p}_c}} 
\newcommand{\sstressy}{{\hat \tau_s}} 
\newcommand{\fstressy}{{\hat \tau_f}} 
\newcommand{\nn}{\nonumber}

\newcommand{\real}{\mathcal{R}} 

\newcommand{\py}{\partial_y}
\newcommand{\pyy}{\partial_{yy}}
\newcommand{\pyyy}{\partial_{yyy}}
\newcommand{\pyyyy}{\partial_{yyyy}}

\numberwithin{equation}{section} 

\begin{document}

\title{Stability of concentrated suspensions under Couette and Poiseuille flow}

\author{Tobias Ahnert%
\thanks{Institut f\"ur Mathematik, Technische Universit\"at Berlin, Stra{\ss}e des 17. Juni 136, 10623 Berlin, Germany}
\and 
Andreas M\"unch%
\thanks{Mathematical Institute, University of Oxford, Andrew Wiles Building, Woodstock Road, Oxford OX2 6GG, UK} 
\and 
Barbara Niethammer%
\thanks{Hausdorff Center for Mathematics, Villa Maria, Endenicher Allee 62, 53115 Bonn, Germany}
\and 
Barbara  Wagner%
\thanks{Weierstrass Institute (WIAS), Mohrenstrasse 39, 10117  Berlin, Germany}
}

\maketitle

\begin{abstract}
The stability of two-dimensional Poiseuille flow and plane Couette flow for concentrated suspensions is investigated. Linear stability analysis of the two-phase flow model for both flow geometries shows the existence of a convectively driven instability with increasing growth rates of the unstable modes as the particle volume fraction of the suspension increases. In addition it is shown that there exists a bound for the particle phase viscosity below which the two-phase flow model may become ill-posed as the particle phase approaches its maximum packing fraction. 
The case of two-dimensional Poiseuille flow gives rise to base state solutions that exhibit a jammed and unyielded region, due to shear-induced migration, as the maximum packing fraction is approached. The stability characteristics of the resulting Bingham-type flow is investigated and connections to the stability problem for the related classical Bingham-flow problem are discussed. 
\end{abstract}

\setcounter{secnumdepth}{3}
\setcounter{tocdepth}{4}

\makeatletter
  \providecommand\@dotsep{5}
\makeatother

\section{Introduction}\label{sec:Intro} 
It is well-known since the work by Orszag \cite{Orszag1971} that two-dimensional Poiseuille flow of Newtonian fluids have a critical Reynolds number of $\mathrm{Re} \approx 5772.22$ beyond which point the flow becomes linearly unstable. The linear stability analysis of parallel shear flows, such as Poiseuille and Couette flow is based on the study of the spectrum of the associated initial boundary value problem for the Orr-Sommerfeld equation, described in detail for example in Drazin and Reid~\cite{Drazin1981}. This analysis does not reveal all the unstable behavior seen in experiments as some nonlinear instabilities do seem to be initiated by linear transient growth of certain modes, which is possible  
since the eigenfunctions of the Orr-Sommerfeld boundary value problem are not orthogonal as discussed in Trefethen et al.~\cite{Trefethen1993}. Some of these modes have time to grow large enough to serve as finite amplitude perturbation and eventually lead to a nonlinear, possibly three-dimensional, instability. 
While the literature on these fundamental hydrodynamic instabilities as well as their route to turbulence is quite extensive, much less is known if non-Newtonian fluids or multiphase liquids are considered \cite{Bolotnov2008,Georgievskii2013,Garifullin1974}.

For the two-phase model equations for concentrated suspensions, which is the focus of this study, it has been shown in Ahnert et al.~\cite{Ahnert2014} that as the maximum packing fraction is approached, plane Poiseuille flow gives rise to jammed and unyielded regions of the solid (particulate) phase. This flow structure is a result of shear-induced migration, a phenomenon first discovered by Leighton and Acrivos \cite{Leighton1987}, and of a yield
stress condition for the solid phase. Hence, understanding the effect of yield stress on the stability properties of the flow is of particular interest.
One of the first studies of the effect of the yield stress on the stability properties can be found in Frigaard et al.~\cite{Frigaard1994}. In their analysis for the plane Poiseuille flow
of a Bingham fluid (one of the simplest cases of a one-phase fluid model with a yield stress) 
they derived a boundary value problem analogon of
the Orr-Sommerfeld problem for Newtonian flow. Further discussions by Frigaard et al.~\cite{Frigaard2003} and more recently by Metivier et al.\ \cite{Metivier2005} and Georgievskii  \cite{Georgievskii2013} showed that the stability properties for plane Poiseuille flow depend critically on the choice of boundary conditions at the yield surface for the associated eigenvalue problem. Using symmetric boundary conditions for the velocity at the yield surface the well-known critical Reynolds number $\RE= 5772.22$ is approached as the Bingham number $\mathrm{B} \to 0$. On the other hand M\'etivier et al.\ \cite{Metivier2005} noted that for their non-symmetric boundary conditions all modes are stable, also as $\mathrm{B} \to 0$. This indicates that the Orr-Sommerfeld-Bingham equation is not a canonical  generalization of the standard Orr-Sommerfeld equation. 

Guided by these investigations, we revisit the formulation of the boundary value problem for the Orr-Sommerfeld-Bingham equations and discuss its implications for the derivation to the eigenvalue problem for the two-phase flow of plane Couette and Poiseuille flow. In particular we show that for the two-phase Poiseuille flow model for concentrated suspensions the conditions at the yield surface of the corresponding eigenvalue problem are non-symmetric.
The stability analysis of the resulting boundary value problem carried out in this study thus constitutes a next step in complexity for the investigation of the dynamical behavior of two-phase flow models with yield-stress. 
The analysis will moreover serve to assess the necessary conditions to address the problem of well-posedness of the two-phase flow model. 

The problem of well-posedness is in fact an inherent property of even the simplest multiphase model equations for suspension flow and many other applications, since its first derivations from an averaging method pioneered by Drew and Passmann~\cite{Drew1971} and Ishii~\cite{Ishii1975}. Nevertheless, such models have found widespread applications and using various forms of regularizations their study started the development of a number of numerical schemes described for example in Stewart and Wendroff~\cite{Stewart1984}. 
The problem of ill-posedness has recently been reviewed by Lhuillier et al.~\cite{Lhuillier2013}. 
In a series of articles, Keyfitz et al.~\cite{Keyfitz1995, Keyfitz2011, Keyfitz2003}
showed for simple cases of two-phase flows that the ill-posedness of the initial boundary value problem is connected to a loss of hyperbolicity in the principal part of the equations. They have begun to generalize the theory for conservation laws in order to connect the arising singular behavior with the existence of a so-called singular shock. The present study is intended to lay the groundwork for future studies concerning the existence of singular shocks in concentrated suspensions. 

After the formulation of the two-phase flow model and the derivation of the eigenvalue problem in Section \ref{sec:GoverningEquations}, our investigations will focus on the stability analysis of the Couette flow problem in Section \ref{sec:Couette}. This problem is instructive since we can simplify the resulting eigenvalue problem considerably and derive criteria for an ill-posedness in the system that is related to the competition between the solid phase viscosity and the collision pressure. The study of these special cases is then used for the design of a reliable numerical scheme for the general eigenvalue problem. 

In addition to the ill-posedness we also find a convection induced instability via a Kelvin-mode ansatz and show that in general, the growth of the unstable mode is transient. However, as the particle volume fraction of the suspension increases the growth rates of the unstable modes increase as well, so that it can become strong enough to possibly trigger finite-amplitude, nonlinear instabilities. 

For the two-dimensional Poiseuille flow, considered in Section \ref{sec:Poiseuille}, simplifications of the resulting eigenvalue problem, that allow analytical work are not possible. Here, our numerical parameter studies show that the ill-posedness as well as the transient growth property occur again, however for different parameter values. 
The main difference to the Couette flow is that for Poiseuille flow there are volume fractions for which unyielded region emerge. The stability of the corresponding yielding surface is the final topic of our investigations. For the derivation of the associated boundary value problem we found it helpful to revisit the formulation of the eigenvalue problem for the Orr-Sommerfeld-Bingham equation. We conclude in Section \ref{sec:Conclusion} with an outlook.

\section{Governing equations for two-phase flow}\label{sec:GoverningEquations}

\subsection{Formulation of the model}\label{subsec:form}

We consider a two-phase flow model of a suspension consisting of solid particles fully dispersed in a liquid medium, that has been derived in Ahnert et al.~\cite{Ahnert2014}. Its derivation is based on an ensemble average process of the incompressible Navier-Stokes equations along the lines of Drew et al.~\cite{Drew1999} with  constitutive laws based on the work by Boyer et al.~\cite{Boyer2011}, that were meant to unify  liquid suspension and granular rheology and enable us to capture the behavior of concentrated suspensions.

In order to state the model, we define some quantities first. Let $\vf_j$ denote the volume fraction of phase $j$, $\vel_j = (\velu_j, \velv_j)$ the velocity, $\prs_j$ the pressure, $\stress_j$ the shear-stress and $\srt_j = \nabla \vel_j + (\nabla \vel_j)^T$ the shear rate, where $j \in \{s,f\}$ and the indices $s$ and $f$ denote the solid or liquid phase, respectively. We use the usual norm
$\|\ts{A}\| = \left(\frac12\cdot \ts{A}:\ts{A} \right)^{\frac12} $
for symmetric tensors. The dimensional model contains the liquid viscosity $\mu_f$, the densities $\rho_j$ and the permeability $K$, for details see  \cite{Ahnert2014}. Using the scales $U_0$ for velocity, $L$ for length as well as $({U_0 \mu_f})/{L}$ for the pressure and the stresses, the governing equations of the two-phase model are
\begin{subequations}\label{eqn:two_phase_governing_eqn}
\begin{align}
 \svf + \fvf &= 1, \label{eqn:two_phase_governing_eqn_a}\\
 \partial_t \vf_f + \nabla \cdot (\vf_f \relax{\vel_f}) &= 0, \label{eqn:two_phase_governing_eqn_b}\\
 \partial_t \vf_s + \nabla \cdot (\vf_s \relax{\vel_s}) &= 0, \label{eqn:two_phase_governing_eqn_c}\\
{\mathrm{Re}} [\partial_t (\vf_f \relax{\vel_f}) + \nabla \cdot (\vf_f \relax{\vel_f} \otimes \relax{\vel_f})]
-\nabla \cdot (\vf_f \relax\relax{\stress_f}) + 
 \vf_f \nabla \prs_f &= -\mathrm{Da}\, \frac{\svf^2}{\fvf}(\fvel - \svel),
\\
\frac{\mathrm{Re}}{r} [\partial_t (\vf_s \relax{\vel_s}) + \nabla \cdot (\vf_s \relax{\vel_s} \otimes \relax{\vel_s})] -
\nabla \cdot (\vf_s \relax\relax{\stress_s}) + 
 \nabla \prs_c + \svf \nabla \prs_f &= \mathrm{Da}\,  \frac{\svf^2}{\fvf}(\fvel - \svel),
\end{align}
\end{subequations}
where the Reynolds number, Darcy's number and the relative density are defined as
\begin{equation}
\mathrm{Re} = \frac{U L \rho_f}{\mu_f}, \qquad
\mathrm{Da} = \frac{L^2}{K}, \qquad
r = \frac{\rho_f}{\rho_s}.
\end{equation}

The non-dimensionalized constitutive laws are a Newtonian stress for the liquid, i.e.
\begin{subequations}
\label{eqn:constitutive_laws}
\begin{align}
 \fstress &= \fsrt.
\end{align}
For the solid phase, either $\|\ssrt\|>0$, then we require
\begin{align}
 {\sstress} &=  \eta_s(\svf) \ssrt, \label{eqn:constitutive_laws:a}\\
 \cprs &=  \eta_n(\svf) \|\ssrt\|,
\end{align}
with
 \begin{align}
   \eta_s(\svf) &= 1+\frac{5}{2} \frac{\svfcrit}{\svfcrit - \svf}
	+ \mu_c(\svf) \frac{\svf}{(\svfcrit - \svf)^2}, 
	\label{eqn:etas_constitutive_law}
	\\
  \mu_c(\svf)  &= \mu_1 + \frac{\mu_2 - \mu_1}{1 + I_0 \svf^2 (\svfcrit - \svf)^{-2}},
\label{eqn:muc_constitutive_law} \\
  \eta_n(\svf) &= \left(\frac{\svf}{\svfcrit - \svf}\right)^2,
  \label{eqn:etan_constitutive_law}
 \end{align}
or $\ssrt=\vc{0}$, and then we let
\begin{align}
	\svf = \svfcrit
\end{align}
and leave $\sstress$ undefined, but impose the inequality
\begin{align}
	\|\sstress\| \leq \mu_1 \cprs. \label{eqn:constitutive_laws:h}
\end{align}
\end{subequations}
The parameters $\mu_1>0$, $\mu_2\geq 0$, $I_0>0$ are material parameters of the friction law for dense suspensions that characterise the effect of particle contacts on the shear viscosity and $\svfcrit$ is the maximum packing fraction which is achieved exactly where the solid jams.
The form of \eqref{eqn:constitutive_laws:a}-\eqref{eqn:constitutive_laws:h} was inferred in the fundamental study \cite{Boyer2011} through a combination of
scaling arguments and careful experiments with a Couette-type flow,
and incorporated in the derivation of a general two-phase flow 
model \cite{Ahnert2014}.
In more detail,  \cite{Boyer2011} observed that the friction in a Couette flow
ofr a dense suspension under a confinement pressure only depends on a single parameter, the ``viscous number'', and that this dependence could be characterised by the volume fraction 
$\phi_{sc}$ when the suspension is at rest, and three additional parameters
$\mu_1$, $\mu_2$, $I_0$. The first of these characterises the leading order behaviour of the friction close to jamming and the others where obtained from the granular flow literature \cite{Cassar2005}. To capture the behaviour for large viscous numbers (large strain rates), their constitutive law also includes Einstein's law for a dilute suspension (the first two terms in \eqref{eqn:etas_constitutive_law}).

The collision pressure field $p_c$ corresponds to the confinement pressure in \cite{Boyer2011} and is an unknown of our governing equations. For the base states
we consider here -- plane Couette and Poiseuille flow -- it turns out to be constant,
with a value that is determined by the solution of the flow problem.

For future reference we note that \eqref{eqn:two_phase_governing_eqn_a}-\eqref{eqn:two_phase_governing_eqn_c}  imply the incompressibility condition 
\be
 \nabla \cdot (\vf_f \relax{\vel_f}+\vf_s \relax{\vel_s}) = 0\, .\label{incompressibility}
\ee

\subsection{Stability problem}\label{subsec:Eigenval}

For the cases of plane Couette flow and two-dimensional Poiseuille flow, stationary solutions of system \eqref{eqn:two_phase_governing_eqn} are derived in \cite{Ahnert2014}. The variables defining these base states depend on $y$ only except for the pressure $\fprsb$, which is a linear function of $x$ only. 
The base state variables are $U_j$, $V_j$, $\Phi_j$, $P_f$, $P_c$  and because $V_j = 0$ for parallel shear flows we obtain 
\begin{align}
\mathbf\Gamma_j &= 
	\begin{pmatrix}
		0 & \partial_y U_j \\
		\partial_y U_j & 0
	\end{pmatrix}, &
\mathbf{T}_f &= 
	\begin{pmatrix}
		0 & \partial_y U_f \\
		\partial_y U_f & 0
	\end{pmatrix}, &
\mathbf{T}_s &= \eta_s(\Phi_s)
	\begin{pmatrix}
		0 & \partial_y U_s \\
		\partial_y U_s & 0
	\end{pmatrix}.
\end{align}
We denote the perturbation variables by lower-case letters with a tilde.
Linearizing about the base states by using the ansatz
\begin{subequations}
\begin{align}
\label{eqn:linearization_ansatz}
\phi_j &= \Phi_j + \delta\tilde\phi_j, &
u_j &=U_j + \delta\tilde{u}_j, &
v_j &= \delta\tilde{v}_j, \\
\dot{\ts{\gamma}}_j &= \mathbf \Gamma_j + \delta\tilde{\dot{\ts{\gamma}}}_j, &
\fprs &= \fprsb + \delta \fprsp, &
\cprs &= \cprsb + \delta \cprsp, \\
\ts{\tau}_j&=\ts{T}_j + \delta \tilde{\ts{\tau}}_j, &&&&
\end{align}
\end{subequations}
where $j \in \{f, s\}$ denote solid and liquid phase and $\delta$ denotes the  small perturbation parameter,  we obtain to order $\delta$ the linearized system
\begin{subequations}
\label{eqn:multiphase_linearized}
\begin{align}
 \fvfp + \svfp &= 0, \\
 \partial_t \fvfp + \partial_x (\fvfb \fvelup + \fvfp \fvelub) + \partial_y (\fvfb \fvelvp ) &= 0, \\
 \partial_t \svfp + \partial_x (\svfb \svelup + \svfp \svelub) + \partial_y (\svfb \svelvp ) &= 0, \\
 \mathrm{Re} [\partial_t (\fvfp \fvelub + \fvfb \fvelup) + \partial_x (2\fvfb \fvelub \fvelup + \fvfp \fvelub^2) &+ \partial_y (\fvfb \fvelub \fvelvp)] - \partial_x (\fvfb \fstressp{}_{11}) \\ \notag
 -\partial_y (\fvfb \fstressp{}_{12} + \fvfp \fstressb_{12}) + \fvfb \partial_x \fprsp + \fvfp \partial_x \fprsb &= -\mathrm{Da} \bigg[\frac{2 \svfb \svfp}{\fvfb} (\fvelub - \svelub) - \\ \notag
 &\frac{\svfb^2}{\fvfb^2}\fvfp (\fvelub - \svelub) + \frac{\svfb^2}{\fvfb} (\fvelup -\svelup)\bigg], \\
 \mathrm{Re} \left[\partial_t (\fvfb \fvelvp) + \partial_x (\fvfb \fvelub \fvelvp)\right] - \partial_x (\fvfb \fstressp{}_{12} + \fvfp &\fstressb_{12}) \\ \notag
 -\partial_y (\fvfb \fstressp{}_{22}) + \fvfb \partial_y \fprsp &= -\mathrm{Da} \bigg[\frac{\svfb^2}{\fvfb} (\fvelvp -\svelvp) \bigg], \\
 \frac{\mathrm{Re}}{r} [\partial_t (\svfp \svelub + \svfb \svelup) + \partial_x (2\svfb \svelub \svelup + \svfp \svelub^2) &+ \partial_y (\svfb \svelub \svelvp)] - \partial_x (\svfb \sstressp{}_{11}) \\ \notag
 -\partial_y (\svfb \sstressp{}_{12} + \svfp \sstressb_{12}) + \partial_x \cprsp + \svfb \partial_x \fprsp + \svfp \partial_x \fprsb &= \mathrm{Da} \bigg[\frac{2 \svfb \svfp}{\fvfb} (\fvelub - \svelub) - \\ \notag
 &\frac{\svfb^2}{\fvfb^2}\fvfp (\fvelub - \svelub) + \frac{\svfb^2}{\fvfb} (\fvelup -\svelup)\bigg], \\
 \frac{\mathrm{Re}}{r} [\partial_t (\svfb \svelvp) + \partial_x (\svfb \svelub \svelvp)] - \partial_x (\svfb \sstressp{}_{12} + \svfp &\sstressb_{12}) \\ \notag
 -\partial_y (\svfb \sstressp{}_{22}) + \partial_y \cprsp + \svfb \partial_y \fprsp &= \mathrm{Da} \bigg[\frac{\svfb^2}{\fvfb} (\fvelvp -\svelvp)\bigg],
\end{align}
\end{subequations}
which is amenable to normal mode analysis and thus we make the ansatz for the perturbation 
\begin{align}
\label{eqn:fourier_ansatz}
\{\tilde\phi_j, \tilde u_j, \tilde v_j, \tilde p_f \} = \{\hat \phi_j(y), \hat u_j(y), \hat v_j(y), \hat p_f(y) \}\, e^{i\alpha x + c t}.
\end{align}
Note that with this choice of ansatz functions an unstable mode fulfills that the real part $\real(c) > 0$. Plugging the ansatz into system \eqref{eqn:multiphase_linearized} yields
\begin{subequations}
\label{eqn:finished_system}
\begin{align}
 -c \svfy + i \alpha (\fvfb \fveluy - \svfy \fvelub) + \partial_y (\fvfb \fvelvy ) &= 0, \label{eqn:finished_system_conti_f} \\
 c \svfy + i \alpha (\svfb \sveluy + \svfy \svelub) + \partial_y (\svfb \svelvy ) &= 0,  \label{eqn:finished_system_conti_s} \\
 \mathrm{Re} [c(-\svfy \fvelub + \fvfb \fveluy) + i \alpha (2 \fvfb \fvelub \fveluy - \svfy \fvelub^2) + \partial_y (\fvfb \fvelub \fvelvy)] \\ \notag
 - i \alpha (\fvfb \fstressy{}_{11}) - \partial_y (\fvfb \fstressy{}_{12} - \svfy \fstressb_{12}) + i \alpha \fvfb \fprsy - \svfy \fprsb_{,x}
  &= - \mathrm{Da} \bigg[\frac{2 \svfb \svfy}{\fvfb} (\fvelub - \svelub) \\ \notag + \frac{\svfb^2}{\fvfb^2} \svfy (\fvelub &- \svelub) + \frac{\svfb^2}{\fvfb}(\fveluy - \sveluy)\bigg], \\
 \mathrm{Re} [c (\fvfb \fvelvy) + i \alpha (\fvfb \fvelub \fvelvy)]
 - i \alpha (\fvfb \fstressy{}_{21} - \svfy \fstressb_{21}) \\ \notag 
 - \partial_y (\fvfb \fstressy{}_{22}) + \fvfb \partial_y \fprsy 
 &= -\mathrm{Da} \bigg[\frac{\svfb^2}{\fvfb}(\fvelvy - \svelvy)\bigg], \\
 \frac{\mathrm{Re}}{r} \Big[c(\svfy \svelub + \svfb \sveluy) + i \alpha (2 \svfb \svelub \sveluy + \svfy \svelub^2)
 + \partial_y (\svfb \svelub \svelvy)\Big] \\\notag
 - i \alpha (\svfb \sstressy{}_{11}) - \partial_y (\svfb \sstressy{}_{12} + \svfy \sstressb_{12})
  + i \alpha \cprsy + i \alpha \svfb \fprsy + \svfy \fprsb_{,x}
  &= \mathrm{Da} \bigg[\frac{2 \svfb \svfy}{\fvfb} (\fvelub - \svelub) \\\notag
  + \frac{\svfb^2}{\fvfb^2} \svfy (\fvelub &- \svelub) + \frac{\svfb^2}{\fvfb}(\fveluy - \sveluy)\bigg], \\
 \frac{\mathrm{Re}}{r} \Big[c (\svfb \svelvy) + i \alpha (\svfb \svelub \svelvy)\Big] - i \alpha (\svfb \sstressy{}_{21} + \svfy \sstressb_{21}) \\ \notag - \partial_y (\svfb \sstressy{}_{22}) + \partial_y \cprsy + \svfb \partial_y \fprsy
 &= \mathrm{Da} \bigg[\frac{\svfb^2}{\fvfb}(\fvelvy - \svelvy)\bigg],
\end{align}
\end{subequations}
with
\begin{subequations}
\begin{align}
\hat \gamma_j &=
 \begin{pmatrix}
 2 i \alpha \jveluy & \partial_y \jveluy + i \alpha \jvelvy \\
 \partial_y \jveluy + i \alpha \jvelvy & 2 \partial_y \jvelvy
 \end{pmatrix}, \\
 \fstressy &= \hat \gamma_f, \\
 \sstressy &= \eta_s'(\svfb) \svfy \Gamma_s + \eta_s(\svfb) \hat \gamma_s, 
 \label{add1}\\
 \cprsy &= \eta_n'(\svfb) \svfy |\Gamma_s| + \eta_n(\svfb) \frac{\partial_y \svelub}{|\partial_y \svelub|} (\partial_y \sveluy + i \alpha \svelvy).
 \label{add2}
\end{align}
\end{subequations}
The primes in \eqref{add1} and \eqref{add2} denote derivatives with respect
to $\svfb$.

We note that this system satisfies a linearized version of the incompressibility condition \eqref{incompressibility}. This leads to spurious eigenvalues in the numerical scheme used to solve the eigenvalue problem. We eliminate these eigenvalues by substitution of the velocity of the liquid phase 
\begin{align}
 \fveluy &= \frac{-1}{i \alpha \fvfb} \left(
 -i \alpha \svfy \fvelub + \partial_y (\fvfb \fvelvy) +
 i \alpha (\svfb \sveluy + \svfy \svelub) + \partial_y (\svfb \svelvy) \right),
 \label{eqn:fluid_uf_elimination}
\end{align}
and the pressure by
\begin{align}
 \label{eqn:final_pf_subsitution}
 \fprsy &= \frac{-1}{i \alpha \fvfb} \bigg( 
 \mathrm{Re} \big[c(-\svfy \fvelub + \fvfb \fveluy) + i \alpha (2 \fvfb \fvelub \fveluy - \svfy \fvelub^2) + \partial_y (\fvfb \fvelub \fvelvy) \big] \\ \notag
 &- i \alpha (\fvfb \fstressy{}_{11}) - \partial_y (\fvfb \fstressy{}_{12} - \svfy \fstressb_{12}) - \svfy \partial_x \fprsb \\ \notag
  &+ \mathrm{Da} \Big[\frac{2 \svfb \svfy}{\fvfb} (\fvelub - \svelub) + \frac{\svfb^2}{\fvfb^2} \svfy (\fvelub - \svelub) + \frac{\svfb^2}{\fvfb}(\fveluy - \sveluy)\Big]
  \bigg).
\end{align}
We note that similar approaches are known from the derivation of the Orr-Sommerfeld equation, where usually the stream function is introduced, which can then be used to eliminate the differential algebraic character from the single phase equations, cf.~\cite{Frigaard1994,Manning2007}.
The remaining equations are
\begin{subequations}
\label{eqn:final_fluid_reduced}
\begin{align}
 c \svfy + i \alpha (\svfb \sveluy + \svfy \svelub) + \partial_y (\svfb \svelvy) &= 0,
 \label{eqn:final_multiphase_conti_s} \\
 \mathrm{Re} [c (\fvfb \fvelvy) + i \alpha (\fvfb \fvelub \fvelvy)] - i \alpha (\fvfb \fstressy{}_{21} - \svfy \fstressb_{21}) \\\notag - \partial_y (\fvfb \fstressy{}_{22}) + \fvfb \partial_y \fprsy
 &= -\mathrm{Da} \frac{\svfb^2}{\fvfb}(\fvelvy - \svelvy), \\
 \frac{\mathrm{Re}}{r} \bigg[c(\svfy \svelub + \svfb \sveluy) + i \alpha (2 \svfb \svelub \sveluy + \svfy \svelub^2) + \partial_y (\svfb \svelub \svelvy) \bigg] \label{eqn:final_multiphase_solid_u}
 \\ \notag
 - i \alpha (\svfb \sstressy{}_{11}) - \partial_y (\svfb \sstressy{}_{12} + \svfy \sstressb_{12}) + i \alpha \hat p_c + i \alpha \fprsy \svfb + \partial_x \fprsb \svfy
  &= \mathrm{Da} \bigg[\frac{2 \svfb \svfy}{\fvfb} (\fvelub - \svelub) \\\notag 
  + \frac{\svfb^2}{\fvfb^2} \svfy (\fvelub &- \svelub) + \frac{\svfb^2}{\fvfb}(\fveluy - \sveluy)\bigg], \\
 \frac{\mathrm{Re}}{r} \bigg[c (\svfb \svelvy) + i \alpha (\svfb \svelub \svelvy)\bigg] - i \alpha (\svfb \sstressy{}_{21} + \svfy \sstressb_{21}) \label{eqn:final_multiphase_solid_v}
 \\\notag 
 - \partial_y (\svfb \sstressy{}_{22}) + \hat \partial_y p_c + \partial_y \fprsy \svfb
 &= \mathrm{Da} \frac{\svfb^2}{\fvfb}(\fvelvy - \svelvy).
\end{align}
\end{subequations}

For the case when the solid phase reaches maximum packing fraction $\svf = \svfcrit$, the momentum equations \eqref{eqn:final_multiphase_solid_u} and \eqref{eqn:final_multiphase_solid_v} lose their validity and condition $\ssrt=\vc{0}$ tells us that the solid phase is confined to rigid motions. Hence, in  this case we drop the two momentum equations and set
\begin{align}
 \svfy &= 0,& 
 \svfb &= \svfcrit,&
 \sveluy &= 0,&
 \svelvy &= 0.
\end{align}
This in turn also eliminates \eqref{eqn:final_multiphase_conti_s} and the  equation for the unyielded region becomes
\begin{align}
\label{eqn:final_solid_reduced}
 \mathrm{Re} [c \fvfb \fvelvy + i \alpha \fvfb \fvelub \fvelvy ] - i \alpha \fvfb \fstressy{}_{21} - \partial_y (\fvfb \fstressy{}_{22}) + \fvfb \partial_y \fprsy = -\mathrm{Da} \frac{\svfb^2}{\fvfb} \fvelvy.
\end{align}
This equation for the unyielded region will only be needed in the Poiseuille flow computation, as the Couette flow does not contain an unyielded region.

\section{Plane Couette flow}\label{sec:Couette}

Consider a planar flow of a fluid confined between two walls at $y = 0$ and $y=L$, where we usually choose $L=1$. The boundary conditions at the lower wall are
\begin{subequations}
\label{eqn:shear_flow_original_bc}
\begin{align}
	\svel = \fvel &= \vc{0} \qquad \text{ at } y = 0,
\end{align}
and for the upper wall are
\begin{align}
	\svel = \fvel &= 
	\begin{pmatrix}
		L \\ 0
	\end{pmatrix} \qquad \text{ at } y = L.
\end{align}
\end{subequations}
System \eqref{eqn:two_phase_governing_eqn} allows the derivation of an explicit solution for the plane Couette flow with base states \cite{Ahnert2014}
\begin{align}
\label{eqn:shear_flow_base}
	\svelub(y) = \fvelub(y) &= y, &
	\fprsb &= C_1, &
	\svfb &= C_2,
\end{align}
where $C_1 \in \mathbb{R}$ and $C_2 \in ]0,\svfcrit[$ are free parameters.

\begin{subequations}
\label{eqn:BCs_shear_case}
Using the boundary conditions \eqref{eqn:shear_flow_original_bc} in our ansatz \eqref{eqn:linearization_ansatz} and \eqref{eqn:fourier_ansatz} yields
\begin{align}
	\sveluy = \svelvy = \fvelvy &= 0 \qquad \text{ at } y = 0 \text{ and } L.
\end{align}
The incompressibility condition \eqref{eqn:fluid_uf_elimination} together with $\svelvy = \fvelvy = 0$ yields
\begin{align}
	\fvfb \partial_y \fvelvy + \svfb \partial_y \svelvy &= 0 \qquad \text{ at } y = 0 \text{ and } L.
\end{align}
\end{subequations}

\subsection{Numerical solution of the spectrum}

We use a finite-difference method for the numerical solution of the system above and use a central scheme of second order for all variables. The pure convection equation of the volume fraction \eqref{eqn:final_multiphase_conti_s} showed an odd-even decoupling, which has been solved using a staggered grid approach. 
The system \eqref{eqn:final_fluid_reduced} with boundary conditions \eqref{eqn:BCs_shear_case}, yielding the generalized eigenvalue problem for $c$, can then be solved using standard solvers. Details of the numerical approximation are given in Appendix \ref{sec:NumericalScheme}. 

\begin{figure}[ht!]
	\includegraphics[width=0.99\linewidth]{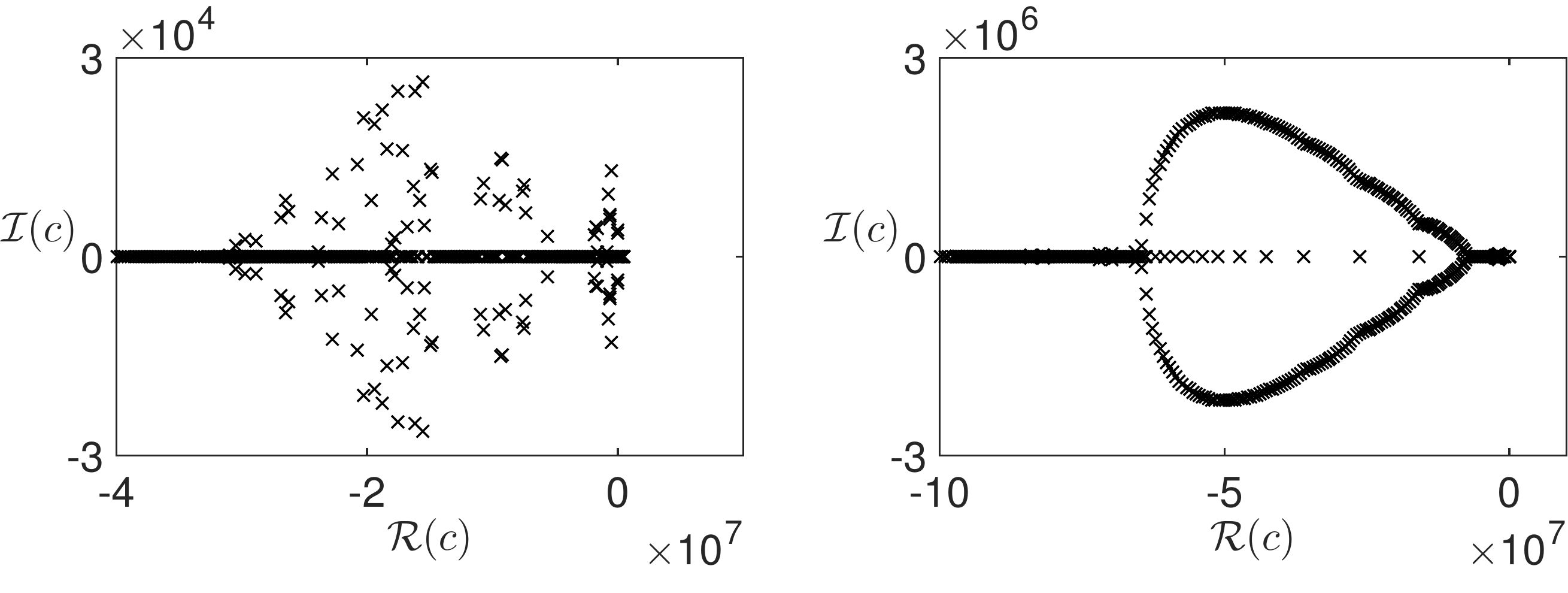}
	\caption{Shown are the two-phase plane Couette flow spectra with parameters chosen as $\RE = 1$, $\DA = 100$, $I_0 = 0.005$, $\mu_2 = \mu_1$, $\svfcrit = 0.63$, $\svfb = 0.99 \svfcrit$, where $\mu_1 = 0.32$ (left) and $\mu_1 = 1$ (right). Both spectra contain unstable eigenvalues near the origin.}
	\label{fig:shear_flow_spectra}
\end{figure}

Compared to the classical problems for the Orr-Sommerfeld equation, the study of the spectrum for our system \eqref{eqn:final_fluid_reduced},  \eqref{eqn:BCs_shear_case} depends on the additional parameters $\DA$, $\mu_1$, $\mu_2$, $I_0$, $\svfcrit$ and $C_2$; $C_1$ drops out of the linearised problem for the Couette flow.
Figure \ref{fig:shear_flow_spectra} shows two spectra for two exemplary choices of parameters, where the parameter values differ in the values of $\mu_1$. One observes that nearly all eigenvalues have negative real parts and, consequently, are stable. 
However, both spectra contain unstable eigenvalues near the origin. 

\begin{figure}[ht!]
	\includegraphics[width=0.99\linewidth]{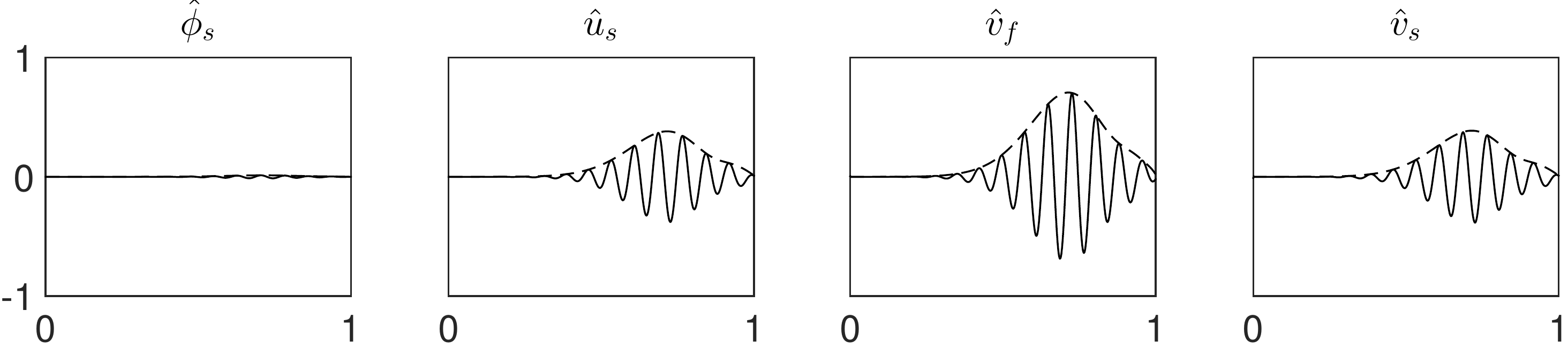}
	\includegraphics[width=0.99\linewidth]{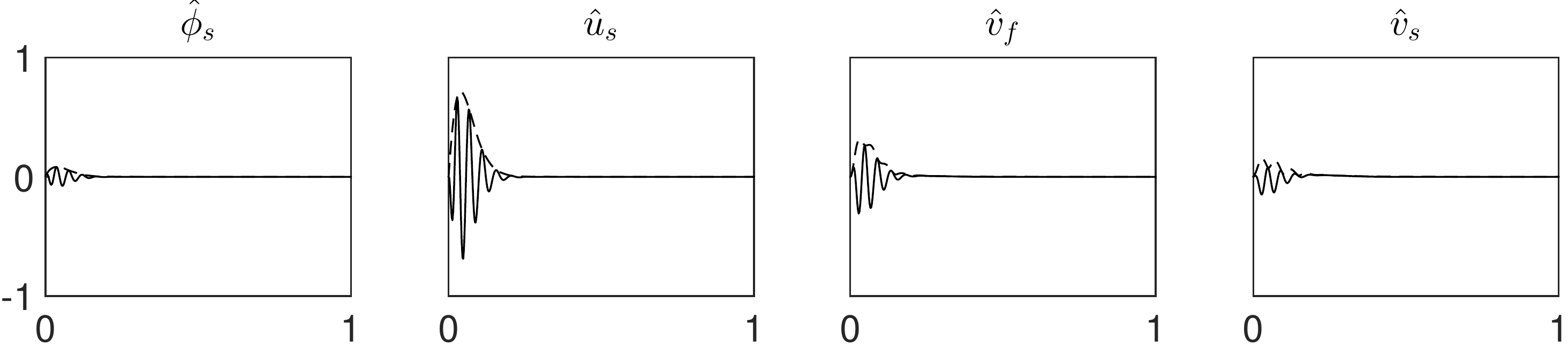}
	\caption{Examples for the two types of unstable mode explained in the text.
	The first is shown in the top row for $\mu_1 = 0.32$, and the second in the bottom row with $\mu_1 = 1$, with the other the parameters as in Figure \ref{fig:shear_flow_spectra}.	}
	\label{fig:shear_flow_eigenmodes_1}
\end{figure}

We could in fact identify two markedly different unstable modes in the system. 
An example of the first is shown in the top row of Figure \ref{fig:shear_flow_eigenmodes_1}. It is observable for $\mu_1 < 1/2$ and the components for $\svfy$ are nearly zero. Most interestingly, as we will show in the following section, the eigenvalues of these modes can grow with $\alpha$ without bounds, which hints at an ill-posedness in the model.
The other unstable mode (bottom row of Figure \ref{fig:shear_flow_eigenmodes_1}) occurs when $C_2$ is close to the maximum packing fraction $\svfcrit$. This mode is concentrated at one end of the interval, and has a visible contribution in $\svfy$. These two modes are analyzed in detail in the following sections.

\subsection{Collision pressure induced ill-posedness}\label{subsec:Collisionpressure}

Our numerical parameter studies  show that the system may lose its well-posedness as soon as  
\begin{align}
\mu_1 < \frac12.
\end{align}
In this case our numerical studies show that the positive real part $\real(c)$  of the eigenvalues grow to infinity as $\svfb \to \svfcrit$ for increasing $\alpha$. As can be seen in Figure \ref{fig:shear_flow_eigenmodes_1} from the corresponding eigenvector, the ill-posedness occurs even for $\svfy = 0$. 
Further, our numerical experiments indicated that the instability mode persists
even upon dropping the quadratic velocity terms 
$\fvfb \fvelub \fveluy$, $\fvfb \fvelub \fvelvy$, $\svfb \svelub \sveluy$ and $\svfb \svelub \svelvy$ in \eqref{eqn:final_pf_subsitution} and \eqref{eqn:final_fluid_reduced}.

These properties can be used to reduce the system \eqref{eqn:multiphase_linearized} further so that we can study and understand the origin of the ill-posedness analytically. Hence, in  \eqref{eqn:multiphase_linearized} we set $\tilde\phi_s = 0$ and neglect the squared velocity parts yielding
\begin{subequations}
\begin{align}
 \partial_x (\svfb \svelup + \fvfb \fvelup) + \partial_y (\svfb \svelvp + \fvfb \fvelvp) &= 0, \\
 \mathrm{Re} \,\partial_t (\fvfb \fvelup) - \partial_x (\fvfb \fstressp{}_{11})
 -\partial_y (\fvfb \fstressp{}_{12}) + \fvfb \partial_x \fprsp  +\mathrm{Da} \left[\frac{\svfb^2}{\fvfb} (\fvelup -\svelup)\right]  &= 0, \\
 \mathrm{Re} \,\partial_t (\fvfb \fvelvp) - \partial_x (\fvfb \fstressp{}_{12})
 -\partial_y (\fvfb \fstressp{}_{22}) + \fvfb \partial_y \fprsp + \mathrm{Da} \left[\frac{\svfb^2}{\fvfb} (\fvelvp -\svelvp)\right]  &= 0, \\
 \frac{\mathrm{Re}}{r} \partial_t (\svfb \svelup) -\partial_x (\svfb \sstressp{}_{11})
 -\partial_y (\svfb \sstressp{}_{12}) + \partial_x \cprsp + \svfb \partial_x \fprsp - \mathrm{Da} \left[\frac{\svfb^2}{\fvfb} (\fvelup -\svelup)\right]  &= 0, \\
 \frac{\mathrm{Re}}{r} \partial_t (\svfb \svelvp) - \partial_x (\svfb \sstressp{}_{12})
 - \partial_y (\svfb \sstressp{}_{22}) + \partial_y \cprsp + \svfb \partial_y \fprsp - \mathrm{Da} \left[\frac{\svfb^2}{\fvfb} (\fvelvp -\svelvp)\right] &= 0.
\end{align}
\end{subequations}
Eliminating the pressure and one of the velocities through the incompressibility conditions, this set of equation allows the standard Fourier ansatz
\begin{align}
	\{\svelup,\svelvp,\fvelvp\} = \{\sveluy,\svelvy,\fvelvy\} e^{i \alpha x + i \beta y + c t},
\end{align}
yielding the $3 \times 3$ matrix system of the form
\begin{align}
	(\ts{A} - c \ts{I}) \vc{u} &= \vc{0},
\end{align}
which is equivalent to
\begin{align}
	\det(\ts{A} - c \ts{I}) &= 0, \qquad \text{ for } \vc{u} \ne \vc{0}, \label{eqn:pc_instability_det_system}
\end{align}
where an instability fulfills $\real(c) > 0$. Equation \eqref{eqn:pc_instability_det_system} is a polynomial of third order in $c$ that can be solved using computer algebra~\cite{Maple}.

The mechanism of the ill-posedness can be observed most clearly in the simple case when $\mathrm{Da} = 0$,  $\alpha = \beta$. For simplicity we also choose $\mathrm{Re} = 1, \mu_2 = \mu_1$, $r = 1$ and drop the $5/2$-term in the viscosity. A closed form solution for the eigenvalues can be derived, which yields the following  amplification factors
\begin{subequations}
\begin{align}
	c_1 &= -2 \alpha^2, \\
	c_2 &= -2 \alpha^2\frac{ (\svfcrit - \svfb)^2 + \mu_1 \svfb}{(\svfcrit - \svfb)^2}, \\
	c_3 &=  2 \alpha^2\frac{(1 - 2\mu_1)\svfb(1-\svfb) -2(\svfcrit - \svfb)^2}{(\svfcrit - \svfb)^2}.
\end{align}
\end{subequations}
It is now easily observed that the amplification factors $c_1$ and $c_2$ are always negative, i.e. are stable and correspond to the liquid and particle viscosity damping, respectively. The third amplification $c_3$ is always negative for $\mu_1 \ge 1/2$, but will always become positive for $\mu_1 < 1/2$ and grows without bound when $\svfb \to \svfcrit$. Hence, the ill-posedness is rooted in a competition between the collision pressure term and the particle viscosity and grows like 
\begin{align}
	c_3 \sim \frac{2\alpha^2}{(\svfcrit-\svfb)^2}.
\end{align}
This eigenvalue grows without bound for increasing $\alpha$ and $\svfb \to \svfcrit$. Thus, it is necessary to set $\mu_1 \ge 1/2$ in order for the problem to be well-posed.

For the general case with Darcy's number set to zero and $\alpha \neq \beta$, the amplification factors are
\begin{subequations}
\begin{align}
	c_1 &= -\frac{\alpha^2+\beta^2}{\RE}, \\
	c_2 &= -r \eta_s \frac{\alpha^2+\beta^2}{\RE}, \\
	c_3 &= 2r \frac{(1-\svfb)(\alpha \beta \eta_n - \svfb \eta_s (\alpha^2 + \beta^2)) - \svfb^2(\alpha^2 + \beta^2)}{\svfb \RE (-\svfb+\svfb r +1)}. \label{eqn:unstable_c_pc_full}
\end{align}
\end{subequations}
Now, the necessary condition for well-posedness is
\begin{equation}
	\alpha \beta \eta_n - \svfb \eta_s (\alpha^2 + \beta^2) \le 0 \text{ for all } \svfb,
\end{equation}
which can be rewritten as
\begin{equation}
	-\frac12 \eta_n\left(\alpha - \beta\right)^2 + \left(\alpha^2 + \beta^2\right) \left(\eta_n - \frac12 \svfb \eta_s\right) \le 0 \text{ for all } \svfb,
	\label{eqn:pc_instability_criterion}
\end{equation}
which shows that the worst case scenario is obtained for $\alpha = \beta$ and gives the necessary criterion, that the particle viscosity must be at least half in size of the collision pressure for all possible choices of parameters. In case of equality $\eta_n = \frac12 \svfb \eta_s$ the mode is stable, since the $-\svfb (\alpha^2 + \beta^2)$ term has a stabilizing influence, which originates from the liquid viscosity. 

For the cases when $\DA > 0$ the eigenmodes are given by 
\begin{subequations}
\begin{align*}
	c_1 &= \frac{1}{2 (\svfb - 1)^2 \RE} \Big(f_1 - (\alpha^2 + \beta^2) (\svfb - 1)^2 (1 + \eta_s r) \\
	&\quad +\sqrt{(\alpha^2+\beta^2)^2 (\svfb - 1)^4 (1 - \eta_s r)^2 - \svfb f_1 - 2 (\alpha^2 + \beta^2) (\svfb - 1)^2 (r \eta_s - 1) f_2}\Big), \\
	c_2 &= \frac{1}{2 (\svfb - 1)^2 \RE} \Big(f_1 - (\alpha^2 + \beta^2) (\svfb - 1)^2 (1 + \eta_s r) \\
		&\quad -\sqrt{(\alpha^2+\beta^2)^2 (\svfb - 1)^4 (1 - \eta_s r)^2 - \svfb f_1 - 2 (\alpha^2 + \beta^2) (\svfb - 1)^2 (r \eta_s - 1) f_2}\Big), \\
	c_3 &= 2r \frac{(1-\svfb)(\alpha \beta \eta_n - \svfb \eta_s (\alpha^2 + \beta^2)) - \svfb^2(\alpha^2 + \beta^2)}{\svfb \RE (1 - \svfb+\svfb r)} - r \frac{\DA\, \svfb}{(\svfb - 1)^2 \RE (1 - \svfb + \svfb r)},
\end{align*}
where
\begin{align*}
	f_1 &= \DA \svfb (\svfb (r-1) - r), \\
	f_2 &= \DA \svfb (\svfb (r+1) - r),
\end{align*}
\end{subequations}
with $f_1<0$ for physically relevant density ratios are between zero and one. 
It shows that the cases $\DA>0$ contain terms that have only a slightly stabilizing effect of order $O(\DA)$, which is not able to compete with the singular terms in $\eta_n$ and $\eta_s$ and thus they do not change the result in an asymptotic sense for $\svfb \to \svfcrit$, unless $\DA$ is artificially chosen to have a specific singular behavior as the maximum packing fraction is approached, see for example \cite{Inkson2014} for recent numerical work on related model equations.

\begin{figure}
	\includegraphics[width=0.9\linewidth]{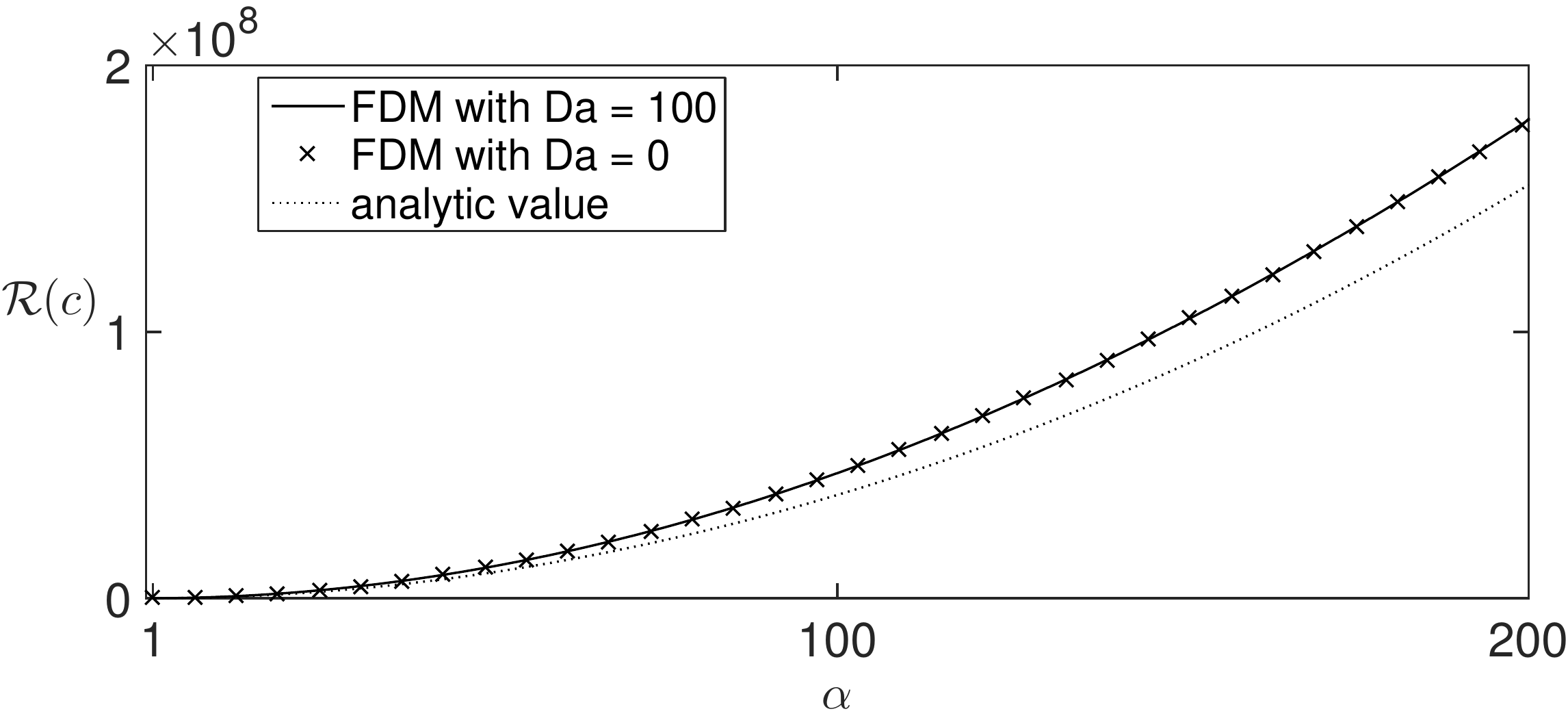}
		\caption{Shown is the dispersion relation of the collision pressure induced ill-posedness for the plane Couette flow with parameters as in Figure \ref{fig:shear_flow_eigenmodes_1}(Top). The analytic curve is computed by equation \eqref{eqn:unstable_c_pc_full}. Comparison of the numerical and the analytical result shows good matching although the numerical simulation uses non-periodic boundary conditions. The curves for different values of $\DA$ are nearly identical, showing the minor influence of the momentum coupling term on the ill-posedness.}
		\label{fig:dispersion_shear_collision}
\end{figure}

Figure \ref{fig:dispersion_shear_collision} shows the singular behavior of the dispersion relation. Comparison between the analytic expression \eqref{eqn:unstable_c_pc_full} and numerical result for different $\DA$ values show good agreement although the numerical results do not use simplifications, e.g. boundary conditions are non-periodic and nonlinear terms are not eliminated in the computations. In particular, the comparison shows that different $\DA$ values hardly change the dispersion curve.

\subsection{Convection induced instability}\label{subsec:Convection}

If $\mu_1 \ge 1/2$ the unstable modes that previously caused the collision pressure induced ill-posedness become stable, however, other unstable modes become apparent. An example of such a mode is shown in Figure \ref{fig:shear_flow_eigenmodes_1} (Bottom). In contrast to the case when $\mu_1<1/2$, the unstable modes in this case have small positive real parts that do not grow with $\alpha$, their modes are non-symmetric and show significant amplifications in $\svfy$. Additionally, if we force $\svfy = 0$ they vanish. Moreover, our parameter studies showed that the instability arises also for vanishing inertial terms. So we set $\RE = 0$ and the linearized system \eqref{eqn:final_fluid_reduced} gives
\begin{subequations}
\begin{align}
 \partial_t \svfp + \svelub \partial_x \svfp + \svfb \partial_x \svelup + \svfb \partial_y \svelvp &= 0, \\
 \partial_x (\fvfb \fvelup) + \partial_y (\fvfb \fvelvp) + \partial_x (\svfb \svelup) + \partial_y (\svfb \svelvp) &= 0, \\
 -\partial_x (\fvfb \fstressp{}_{11}) 
 -\partial_y (\fvfb \fstressp{}_{12} + \fvfp \fstressb_{12})
 + &\fvfb \partial_x \fprsp = \\ \notag -\mathrm{Da} \bigg[\frac{2 \svfb \svfp}{\fvfb} (\fvelub - \svelub) - 
 &\frac{\svfb^2}{\fvfb^2}\fvfp (\fvelub - \svelub) + \frac{\svfb^2}{\fvfb} (\fvelup -\svelup)\bigg], \\
 - \partial_x (\fvfb \fstressp{}_{12} + \fvfp \fstressb_{12})
 - \partial_y (\fvfb \fstressp{}_{22}) + \fvfb \partial_y \fprsp &= -\mathrm{Da} \bigg[\frac{\svfb^2}{\fvfb} (\fvelvp -\svelvp)\bigg], \\
 -\partial_x (\svfb \sstressp{}_{11}) 
 -\partial_y (\svfb \sstressp{}_{12} + \svfp \sstressb_{12}) + &\partial_x \tilde \cprs + \svfb \partial_x \fprsp = \\ \notag
 \mathrm{Da} \bigg[\frac{2 \svfb \svfp}{\fvfb} (\fvelub - \svelub) -
 \frac{\svfb^2}{\fvfb^2}&\fvfp (\fvelub - \svelub) + \frac{\svfb^2}{\fvfb} (\fvelup -\svelup)\bigg], \\
 - \partial_x (\svfb \sstressp{}_{12} + \svfp \sstressb_{12})
 - \partial_y (\svfb \sstressp{}_{22}) + \partial_y \tilde \cprs + \svfb &\partial_y \fprsp = \mathrm{Da} \bigg[\frac{\svfb^2}{\fvfb} (\fvelvp -\svelvp)\bigg].
\end{align}
\end{subequations}
A direct use of the Fourier ansatz is not helpful for this system, as the convective term $\svelub \svfp{}_{,x}$ would introduce derivatives in the wave-number $\alpha$. However, the base state $\svelub = \fvelub = y$ makes it suitable for a Kelvin-mode ansatz \cite{Tatsuno2001}, which consists of two steps - firstly, using the method of characteristics and, secondly, using a Fourier transformation. The method of characteristics eliminates the convective part, but introduces time dependencies in previously stationary parts of the equation. Eventually, the spatial coordinates of the system are transformed into Fourier modes, yielding an ordinary differential equation in time, that can be studied in order to understand the stability properties of the original system.

Therefore, we first use the transformation
\begin{align}
\xi = x - y t\quad\mbox{and}\quad \mathsf{y} = y,
\end{align}
followed by a Fourier ansatz in space only, that is 
\begin{align}
\label{eqn:kelvin_mode_ansatz}
\{\svfp, \svelup, \svelvp, \fvelvp \} = \{\svfy(t) ,\sveluy(t), \svelvy(t), \fvelvy(t) \}e^{i \alpha \xi + i \beta \mathsf{y}},
\end{align}
which gives the system
\begin{subequations}
\begin{align}
	0 = \partial_t \svfy + \svfb ((i\beta - t i \alpha)\svelvy + i \alpha &\sveluy), \\
	\fveluy = \frac{-1}{i \alpha \fvfb}(i \alpha \svfb \sveluy + (i \beta - t i \alpha) (&\fvfb \fvelvy + \svfb \svelvy)), \\
	\fprsy = \frac{-1}{i \alpha \fvfb}(2 \alpha^2 \fvfb \fveluy - (i \beta - t i \alpha)&(\fvfb ((i\beta - t i \alpha)\fveluy + i \alpha \fvelvy) - \svfy)+ \mathrm{Da}\frac{\svfb^2}{\fvfb}(\fveluy - \sveluy)), \\
	-i \alpha (\fvfb ((i\beta - t i \alpha)\fveluy + i \alpha \fvelvy) - &\svfy) - 2 \fvfb (i\beta - t i \alpha)^2 \fvelvy \\
	& + \fvfb (i\beta - t i \alpha) \fprsy
	+ \mathrm{Da} \frac{\svfb^2}{\fvfb}(\fvelvy - \svelvy) = 0, \notag \\
	-i \alpha \svfb \eta_s 2 i \alpha \sveluy  - (i\beta - t i \alpha) (\svfb \eta_s &((i\beta - t i \alpha)\sveluy + i \alpha \svelvy) + \svfb \eta_s' \svfy + \svfy \eta_s) + i \alpha \cprs \\
	&+ i \alpha \svfb \fprsy 
	- \mathrm{Da} \frac{\svfb^2}{\fvfb}(\fveluy - \sveluy) = 0, \notag \\
	-i \alpha (\svfb \eta_s ((i \beta - t i \alpha)\sveluy + i \alpha \svelvy) + &\svfb \eta_s' \svfy + \svfy \eta_s) - 2 \svfb \eta_s (i\beta - t i \alpha)^2 \svelvy \\
	+ \svfb (i\beta - t i \alpha) &\fprsy 
	- \mathrm{Da} \frac{\svfb^2}{\fvfb} (\fvelvy - \svelvy) + (i \beta - t i \alpha) \cprs = 0. \notag
\end{align}
\end{subequations}
This is of the form
\begin{align}
	\begin{pmatrix}
	A_{11} & A_{12} \\ A_{21} & A_{22}
	\end{pmatrix}
	\begin{pmatrix}
	\svfy \\
	\vel
	\end{pmatrix}
	= 
	\begin{pmatrix}
		-\svfy{}_{,t} \\
		\vc{0}
	\end{pmatrix}.
\end{align}
Thus, using the negative Schur complement $S = -(A_{11} - A_{12} A_{22}^{-1} A_{21})$ of $A_{22}$ we get the ordinary differential equation
\begin{align}
	\svfy{}_{,t}(t) = S(t) \svfy(t), \label{eqn:conv_inst_ode}
\end{align}
with solution to \eqref{eqn:conv_inst_ode} 
\begin{align}
	\svfy(t) = \svfy{}(0) \cdot e^{\int_0^t S(T) \ud T}, \label{eqn:conv_inst_ode_solution}
\end{align}
so we expect a perturbation to grow for times $t$ with $\real(S(t)) > 0$ and to shrink for $\real(S(t)) < 0$.

Interestingly, it is possible to obtain analytic expressions for $S$ for special cases. If we set $\mathrm{Da} = 0$ and denote $f_1=\svfb-1$, $f_2=\alpha^2+f_3^2$ and  $f_3 = \beta - t \alpha$, then using computer algebra~\cite{Maple}, we obtain
\begin{align}
S &= 
\frac{f_1\Big[
\eta_n (\eta_s + \eta_s' \svfb) (\alpha^2 - f_3^2)^2 + \eta_s \svfb f_2 [2 \alpha f_3 (\eta_s + \eta_s' \svfb) - \eta_n' f_2] \Big] - 2 \eta_s \svfb^2 f_2 \alpha f_3
}
{
2 \eta_s f_2 \Big[f_1 \left(\svfb \eta_s f_2-\eta_n \alpha f_3  \right) -  f_2 \svfb^2 \Big]
}.
\label{eqn:S_def}
\end{align}
From a theoretical point of view, the Kelvin-mode ansatz first transforms a non-Hermitian differential operator into a Hermitian operator, which allows for a spectral analysis. By the spectral theorem a Hermitian operator has only real eigenvalues, the eigenfunctions are orthogonal and form a complete set. 
Hence, the Schur complement $S$ is always real and combinations of modes $\alpha$ and $\beta$ only occur in even orders. Contrary to the analytic approach, the numerical eigenvalues computed by the full problem posses nonzero imaginary parts.

\begin{figure}[ht!]
\includegraphics[width=0.99\linewidth]{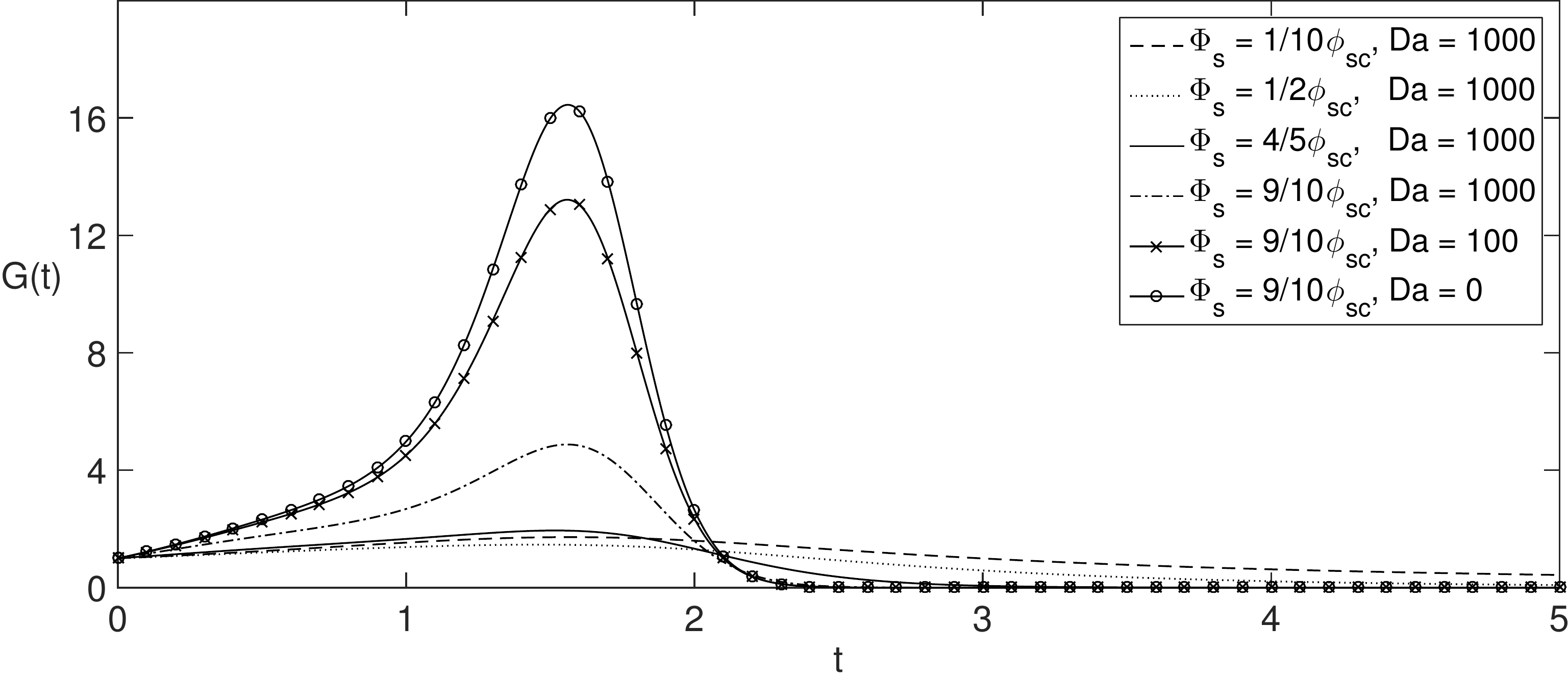}
	\caption{Growth factor for a typical parameter choice of $\alpha = 5, \beta = 8, \svfcrit = 0.63, \mu_1 = \mu_2 = 1, \RE = 0$ and different solid volume fractions and Darcy's numbers. The transient growth behavior can obtain huge values, depending on how close $\svfb$ is to the maximum packing value. For the stated constitutive laws of $\eta_n$ and $\eta_s$ and for long times $t$ the growth is always damped, i.e. $G \to 0$ for $t \to \infty$. Nonzero Darcy's numbers have a stabilizing effect, but do not eliminate the instability completely.}
	\label{fig:growth_factor}
\end{figure}

As one is interested in the growth of an initial perturbation $\svfy{}(0)$, it is conventional to discuss the growth factor defined as \cite{Schmid1994, Schmid2007}
\begin{align}
	G(t) = \sup_{\svfy(0) \ne 0} \left|\frac{\svfy(t)}{\svfy(0)}\right| = \left|e^{\int_0^t S(T) \ud T}\right|.
\end{align}
Figure \ref{fig:growth_factor} shows the typical behavior of the growth factor for a range of parameter choices. 

Moreover, the long time limit of $S$ with the constitutive laws \eqref{eqn:constitutive_laws} and $\mu_1 = \mu_2$ can be computed as
\begin{align*}
	\lim_{t \to \infty} S = 
	\frac{(1 - \svfb) \svfb (7 \svfcrit^2 - 2 \svfb^2)}
	{[2 \svfb (\mu_1 + \svfb) - 9 \svfb \svfcrit + 7 \svfcrit^2] [2 \mu_1 (\svfb-1) \svfb - (\svfb - \svfcrit) (-7 \svfcrit + \svfb (2 + 5 \svfcrit))]}.
\end{align*}
This expression is negative as long as $0 < \svfb < \svfcrit$ and zero for $\svfb \in \{0, \svfcrit\}$, which shows the growth factor $G$ always becomes zero for $t \to \infty$. The expression for $\mu_1 \ne \mu_2$ is more involved, but contains the same behavior. Thus, for all other parameters fixed and $t \to \infty$, the value of $S$ becomes always negative for our constitutive laws \eqref{eqn:constitutive_laws}.

Yet, this convergence is not uniform in $\alpha$ and $\beta$ because using the transformation $\beta = C_1 \alpha$ with $C_1 \in \mathbb{R}$, the Schur complement becomes
\begin{align}
\label{eqn:S_C1_ratio}
S = \frac{f_1[\eta_n (\eta_s + \eta_s' \svfb) (1 - \tilde f_3^2)^2 + 
 \eta_s \svfb \tilde f_2 (2 \tilde f_3 (\eta_s + \eta_s' \svfb) - \eta_n' \tilde f_2)]
  - 2 \eta_s \svfb^2 \tilde f_2 \tilde f_3}
 {2 \eta_s \tilde f_2 (f_1 (\svfb \eta_s \tilde f_2 - \eta_n \tilde f_3 ) - \tilde f_2 \svfb^2)},
\end{align}
where $\tilde f_2 = 1 + \tilde f_3^2$ and $\tilde f_3 = C_1 - t$, which is independent of $\beta$ and $\alpha$. Thus, only the mode ratio $C_1$ is of significance for the damping of a perturbation. 

\subsubsection*{Remark}
We note that this observation may point to a process that transforms the transient into infinite growth. 
It is well-known that nonlinearities transport perturbations from one mode to another, see e.g.~\cite{Pope2000}. This process is generally referred to as energy cascade \cite{Pope2000} and is also known to occur in multiphase models \cite{Bolotnov2008}. Thus, a perturbation being transported to bigger ratios, such that $\tilde f_3$ stays constant over time, can grow infinitely large in magnitude. 
In order for $\tilde f_3$ to stay constant the ratio $C_1$ must grow linear in time, which requires a change of frequency of the perturbation. This means an observable instability might shift its Fourier modes from low to high frequencies over time, which is a mechanism able to produce shocks as is known from the inviscid Burgers equation~\cite{Muraki2007}. Alternatively to a creation of a shock, the highest frequencies might be damped by another nonlinear effect, which in turn might result in a turbulent behavior, that transports perturbations into smaller structures, which are being damped when they approach a critical length scale \cite{Pope2000}. This would correspond to the well-known Kolmogorov's hypothesis for single phase media~\cite{Pope2000}.

\subsubsection{Comparison with the full system} 
In order to understand the stability behavior of the full system, we have to understand the connection between the growth factor $S(t)$ and the unstable modes seen in the finite-difference approximation of the full system, considered in their appropriate spaces. 

$S$ depends on the Fourier modes $\alpha, \beta$ and on time $t$, whereas the finite-difference numerical approximation depends on the Fourier modes $\alpha$, $c$ and the spatial variable $y$. Considering the frozen system at $t=0$, we would have a constant growth $c = S(0)$. This in turn together with equation \eqref{eqn:conv_inst_ode_solution} implies our growth is of the form
\begin{align}
	\svfy &= \svfy(0) \mathrm{e}^{c t},
\end{align}
but this and equation \eqref{eqn:kelvin_mode_ansatz} implies
\begin{align}
	\svfp &= \svfy(0) \mathrm{e}^{c t + i\alpha x + i \beta y}. \label{eqn:kelvin_mode_freeze_0}
\end{align}

Now, the ansatz for the FDM is
\begin{align}
	\svfp &= \svfy(y) \mathrm{e}^{c t + i\alpha x}.
\end{align}
Suppose $\svfy(y)$ is a periodic function, then rewriting $\svfy(y)$ as a Fourier series on a domain $[0,L]$ yields
\begin{align}
	\svfp &= \sum_{k = -\infty}^{\infty} \widehat \svf(k) \mathrm{e}^{c t + i\alpha x + i y 2\pi k /L}, \label{eqn:fdm_fourier_representation}
\end{align}
where $\widehat \svf(k)$ represents the $k$-th Fourier coefficient.
Comparison of \eqref{eqn:kelvin_mode_freeze_0} and \eqref{eqn:fdm_fourier_representation} shows, that our FDM computes the frequencies
\begin{align}
	\beta = \frac{2\pi k}{L},
\end{align}
with $k \in \mathbb{Z}$ and $L$ the domain size. In order for a direct comparison to work, we therefore need to change the boundary conditions \eqref{eqn:BCs_shear_case} to periodic boundary conditions and have to consider small domain sizes $L$. For large $L$ the non-periodic base state $\svelub = y$ has a dominant influence on the solution, which makes a direct comparison of the non-periodic numeric and periodic analytic results impossible.
If the non-periodicity becomes dominant we do not see single frequencies, but rather a sum of several modes next to the boundaries, which always occur in pairs - one on each wall - see Figure \ref{fig:shear_flow_eigenmodes_1} (Bottom). In this case the real part of the maximum amplification is always smaller than $S(0)$, hinting at a damping effect of the boundary.

If we set the collision pressure to zero and use Newtonian viscosity, i.e. $\eta_n = \eta_n' = \eta_s' = 0$ and $\eta_s = 1$, then we still get $S > 0$ for some time. Hence,  this instability is not driven by a collision pressure or a viscosity driven effect, but rather caused by the convection of the flow.

Analytic results for nonzero Darcy's number could not be derived. Nevertheless, numerical solutions for $\DA>0$ showed the momentum coupling term has a stabilizing effect, but is not capable to completely eliminate this instability. Even for very large Darcy's numbers, i.e.~$\DA > 10000$, a small transient growth is observable, cf.~Figure \ref{fig:growth_factor}.

\subsubsection*{Remark} 
A possible physical explanation of the instability is a resistance to high volume fractions in the model. For fluid region with near maximum packing a small perturbation is enough to disperse the densely packed particles. However, this instability is of a highly nonlinear nature for $\svfb \approx \svfcrit$, as a small change in $\svfb$ induces a large change in viscosity and particle pressure.

\section{Poiseuille flow}\label{sec:Poiseuille}

Two-dimensional Poiseuille flow is another seemingly simple example for a fluid flow. However, in contrast to Couette flow, it contains four major complications. First, the base state is not given in closed form anymore, so a stability analysis is much harder. Second, it does contain a plug-flow region, where the linearized set of equations change. Third, the conditions at the yield surface are non-trivial and are derived here explicitly. Last, the well-known loss-of-hyperbolicity problem \cite{Lhuillier2013,Keyfitz2003} that is connected to the ill-posedness, enters as soon as the velocities of the solid and liquid phases are different, which is the case for Poiseuille, but not for Couette flow.

\subsection{Bingham flow revisited}\label{subsec:Bingham}

One of the signatures of our two-phase flow model is that it contains a yield-stress similar to the classical (single-phase) Bingham fluid. Moreover, the stability properties of the Poiseuille flow of a Bingham fluid is a well-studied and intensely analyzed problem, see the review by Frigaard et al.\cite{Frigaard2003} and the discussion in \cite{Frigaard1994, Metivier2005, Pavlov1974}.
In addition, our derivation of the yield-surface boundary conditions of the two-phase model is guided by the derivation for the classical Bingham model. 

It is therefore instructive to revisit the problem of Poiseuille flow for a Bingham fluid, in particular to specify and motivate the yield-surface conditions for the stability problem in the two-phase flow case.

Let us consider the governing equation for the Bingham flow, which are the Navier-Stokes equations with a yield-stress constitutive law \cite{Frigaard1994}, i.e.
\begin{subequations}\label{eqn:Bingham_conti}
\begin{align}
	\nabla \cdot \vel &= 0, \\
	\partial_t\vel + (\vel \cdot \nabla)\vel &= \nabla \cdot \left(\stress - \prs \ts{I}\right),
\end{align}
with
\begin{align}
	\stress &= \frac1\RE \left(1 + \frac{B}{|\srt|}\right) \srt &\text{ for } |\stress| &\ge B/\RE, \label{cccc} \\
	\srt &= \ts{0} &\text{ for } |\stress| &< B/\RE. \label{dddd}
\end{align}
\end{subequations}
The boundary conditions for Poiseuille flow are the no-slip boundary conditions
\begin{align}
	\vel &= \vc{0} \qquad \text{ at } y \in \{-1,1\}
\end{align}
and
continuity of the velocity and normal shear rates at the yield-surface
\cite{Frigaard1994,Frigaard2003,Thual2010,Metivier2005,Huilgol2015}
\begin{align}\label{byc}
	\jumpop{\vel} &=0, \qquad  \jumpop{\srt \cdot \vc{n}} = \vc{0} \qquad \text{ at } y = \pm y_B.
 \end{align}
The second of the two equations in \eqref{byc} 
can in fact be inferred from the condition $|\stress|=B/\RE$ at the yield 
surface \cite{Metivier2005} but is more useful for the linear stability analysis
in this form.

These equations have been non-dimensionalized by scaling the length by $2L$, the velocity by $U_0$, the time by $2L/U_0$ and stress by $\rho U_0^2$, which
introduces the Reynolds number $\RE = \rho U_0 L / \mu_0$ and the Bingham number $\mathrm{B} = {\stresscmpt_0 L}/({\mu_0 U_0})$, where $\rho$, $\mu_0$ and $\stresscmpt_0$ denote the density, viscosity and yield-stress, respectively.
Then, making the assumption of independence of time $t$ and streamwise direction $x$ of the velocities and stress, one can derive the non-dimensionalized base state \cite{Frigaard1994}
\begin{align}
\velb = 
	\begin{cases}
		1, &\text{ for } 0 \le |y| < y_B \\
		1 - \left(\frac{|y| - y_B}{1/2 - y_B}\right)^2, &\text{ for } y_B \le |y| \le 1/2
	\end{cases},
\end{align}
where $y_B = -\frac{B}{\RE P}$ and $P < 0$ is the pressure gradient.
Using linearization and a normal-mode ansatz, one derives the Orr-Sommerfeld-Bingham equation for the mode $\hat{v}$, cf.~\cite{Frigaard1994}
\begin{subequations}
\label{eqn:Bingham}
\begin{align}
	i \alpha \mathrm{Re} \left((\velb - c)\left(\pyy \hat{v} - \alpha^2 \hat{v}\right) - \hat{v} \pyy\velb \right) = \pyyyy \hat{v} - 2 \alpha^2 \pyy \hat{v} + \alpha^4 \hat{v} - 4 \alpha^2 \mathrm{B} \py\left(\frac{\py \hat{v}}{|\py\velb|}\right),
\end{align}
with boundary conditions
\label{eqn:Bingham_BCs}
\begin{align}
	\hat{v} = \py \hat{v} &= 0 &\text{ at } y &= \pm 1/2, \\
	\hat{v} = \py \hat{v} &= 0 &\text{ at } y &= \pm y_B, \\
	\pyy \hat{v}  &= \pm \frac{-2 i \alpha h}{(1/2 - y_B)^2} &\text{ at } y &= \pm y_B.
\end{align}
\end{subequations}
For a derivation of the base state, the Orr-Sommerfeld-Bingham equation and the boundary conditions see Appendix \ref{sec:BinghamBCs}.

The boundary value problem \eqref{eqn:Bingham} has been implemented using a finite difference method with a central scheme, see Appendix \ref{sec:NumericalScheme} for details on the scheme. Since the problem contains a singularity at the yield-surface $y=y_B$, we also implemented a shooting method with Riccati transformation as used in \cite{Frigaard1994}. Both methods gave accurate results, but the finite difference method creates a generalized eigenvalue problem, that can be solved with the help of standard solvers, giving the whole discrete spectrum at once. While the shooting method avoids spurious eigenmodes, it is much harder to find all the relevant eigenmodes.

We note first that for the range of values of $\mathrm{B}$, $\mathrm{Re}$ and $\alpha$ discussed in the literature, no unstable mode was found, in agreement with M\'etivier et al.\ \cite{Metivier2005}. 
However, inspired by the analysis of the Orr-Sommerfeld system \cite{Orszag1971}, the symmetric boundary condition $\py\velvy =0= \pyyy\velvy $ has also been studied by 
Frigaard et al.\ \cite{Frigaard1994}.
Using these symmetric boundary conditions the well-known critical Reynolds number $\RE = 5772.22$ is approached as $\mathrm{B} \to 0$, while for the boundary conditions \eqref{eqn:Bingham_BCs} all modes are stable also as $\mathrm{B} \to 0$, as noted by M\'etivier et al.\ \cite{Metivier2005} which shows that the Orr-Sommerfeld-Bingham equation is not a canonical generalization of the standard Orr-Sommerfeld equation.

\begin{figure}[ht!]
	\includegraphics[width=0.49\linewidth]{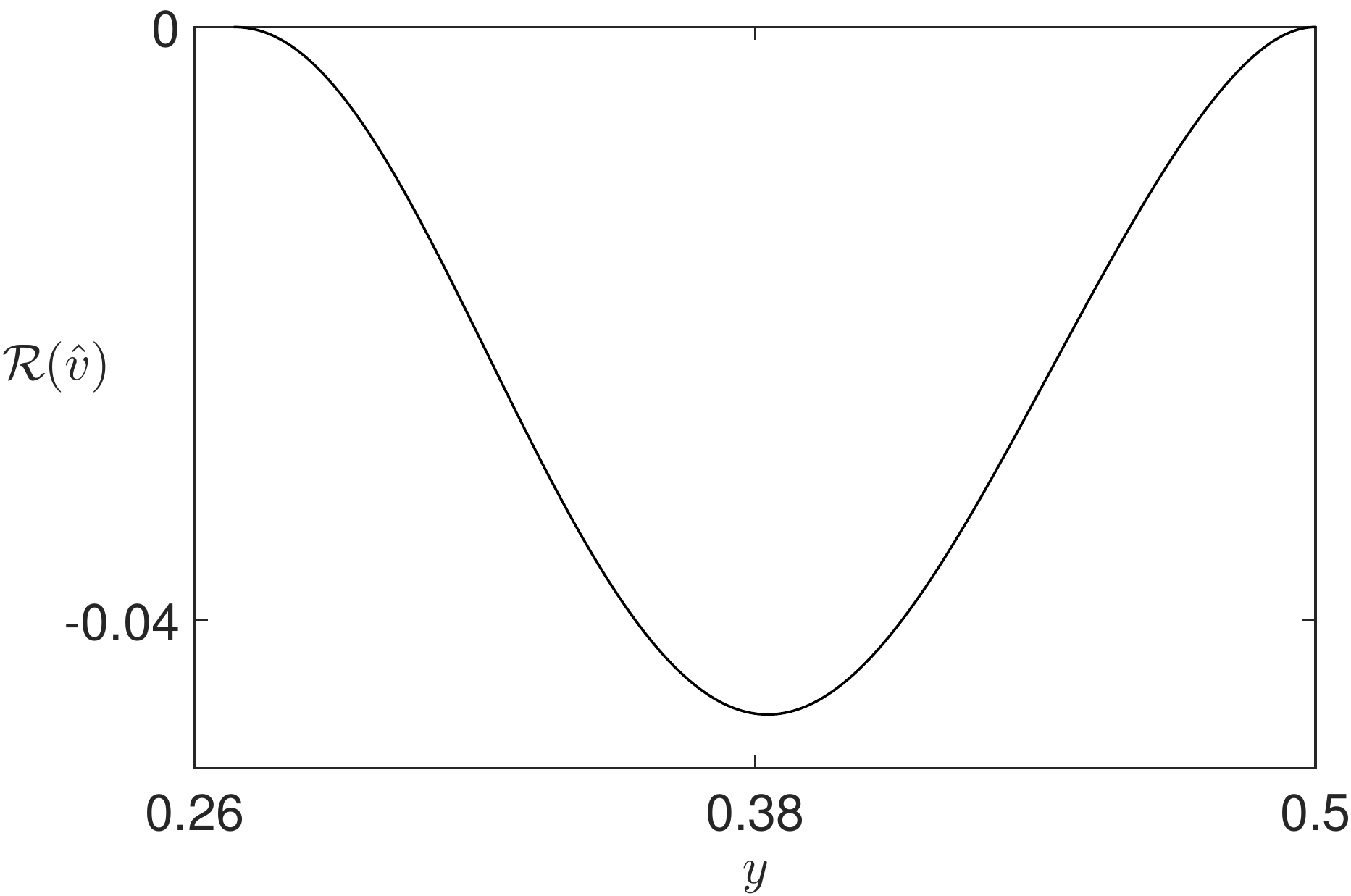}
	\includegraphics[width=0.49\linewidth]{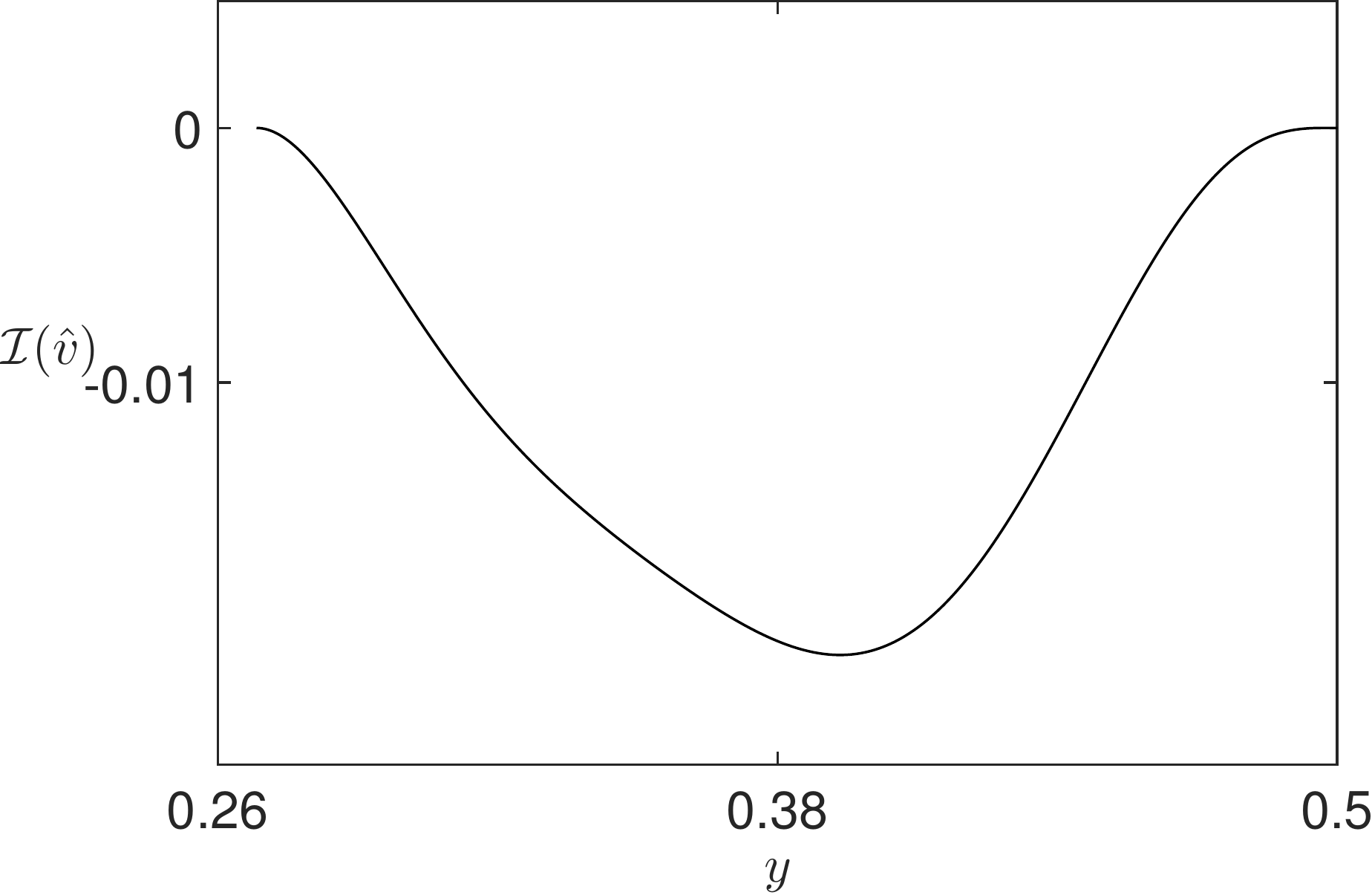} \\
	\includegraphics[width=0.49\linewidth]{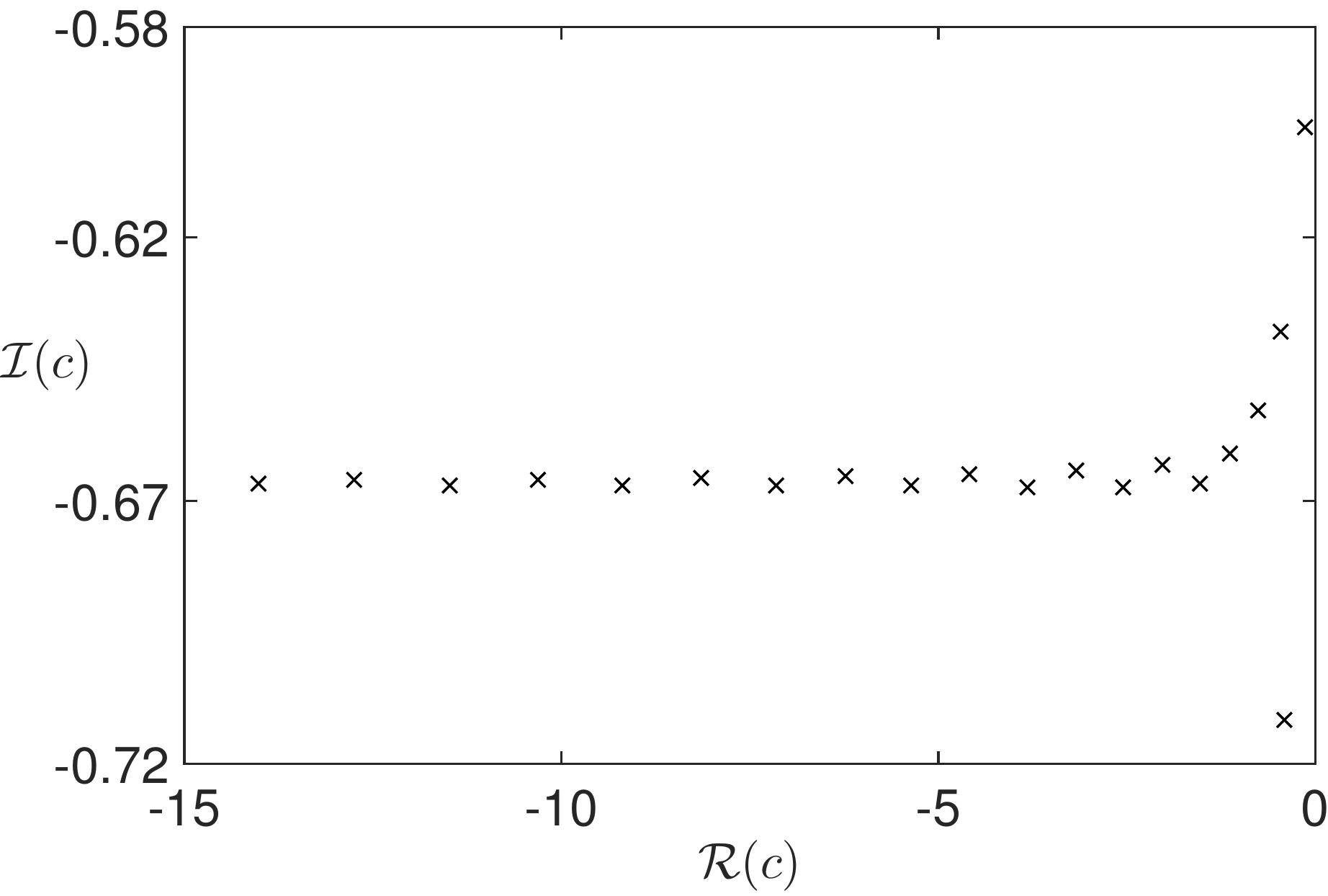}
	\includegraphics[width=0.49\linewidth]{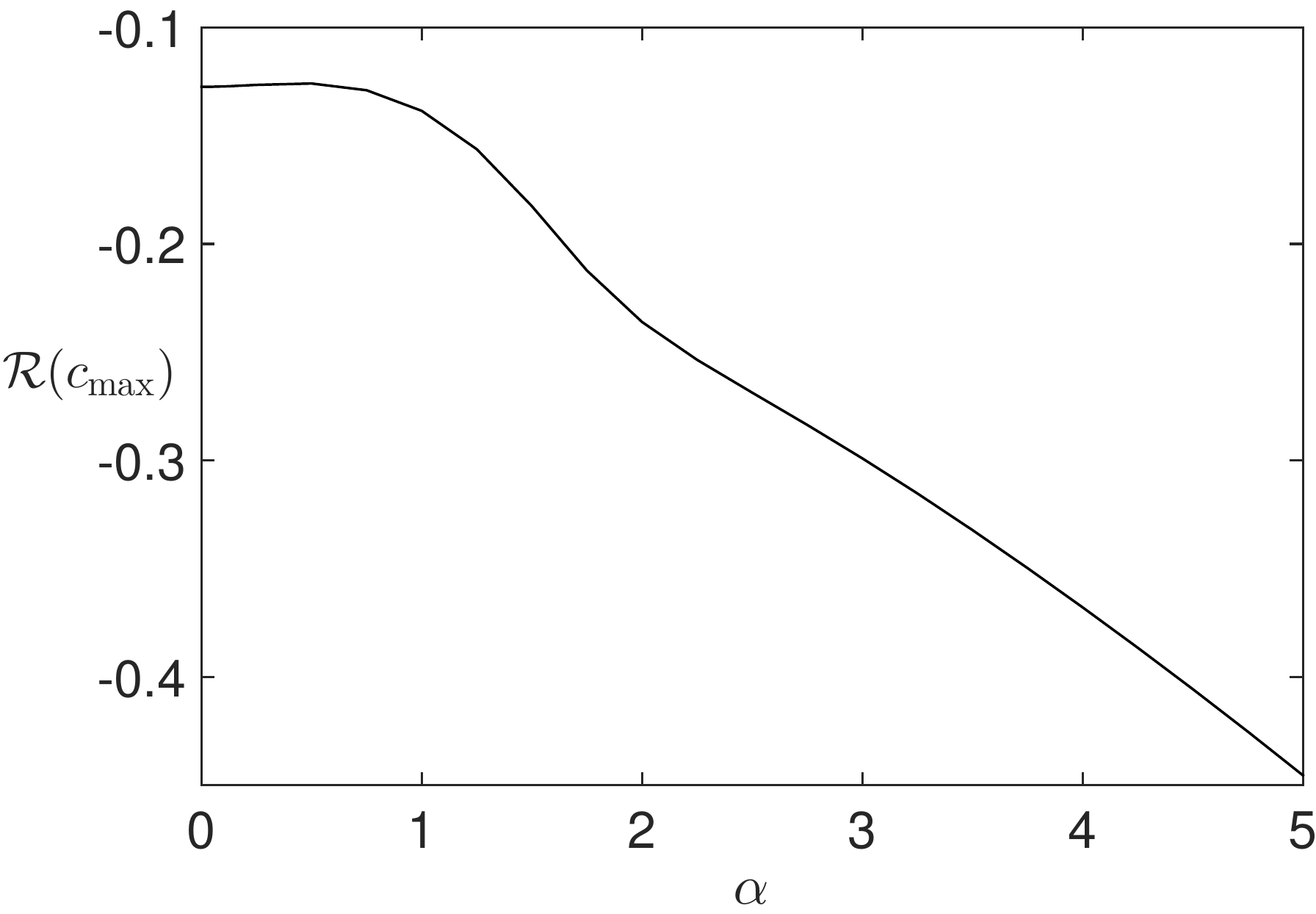}
	\caption{Shown is the real and imaginary part of the most unstable mode for $\mathrm{B} = 10$, $\mathrm{Re} = 5772.22$ and $\alpha = 1$ (top row), for $y$ between $Y_B$ and $1/2$. Also shown is the part of the spectrum with the most unstable modes (bottom, left side) and the dispersion relation of the most unstable mode (bottom, right side).}
	\label{fig:single_phase_results}
\end{figure}

Figure \ref{fig:single_phase_results} shows the results for the classical Bingham model. As can be seen from the spectrum, no eigenvalue has a positive real part, thus the model is linearly stable.

\subsection{Two-phase flow model}
\label{subsec:TwoPhaseFlowModel}

\subsubsection{Base state}
\label{subsec:Basestate}
The Poiseuille flow ansatz is to consider a stationary problem with no-slip boundary conditions
\begin{align}
	\svel = \fvel &= \vc{0} \qquad \text{ at } y = \pm 1/2,
\end{align}
where all quantities, except for the pressure depend only on $y$, i.e.
\begin{align}
	\fvf  &= \fvf(y), & 
	\svf  &= \svf(y), &
	\fvel &= \fvel(y), &
	\svel &= \svel(y), &
	\fprs &= \fprs(x,y),
\end{align}
and, for simplicity, demand the solution to have exactly one plug-flow for $0 \le |y| \le y_B$. At the yield-surface, we demand continuity of the solid and liquid velocities and the normal shear rates similar to the Bingham flow case, i.e.
\begin{align}
	\jumpop{\svel} = \jumpop{\fvel} = \jumpop{\ssrt \cdot \vc{n}} = \jumpop{\fsrt \cdot \vc{n}} &= 0 \qquad \text{ at } y = \pm y_B. \label{eqn:multiphase_yield_surface_conditions}
\end{align}
Note, we did not assume continuity of the tangential shear rates or solid volume fraction, since this would overdetermine the system. For parallel shear flows conditions \eqref{eqn:multiphase_yield_surface_conditions} imply these continuities, which is used in the derivation of the base states.
The base state for the two-phase model has been derived in \cite{Ahnert2014} and it yields a linear liquid pressure $\fprsb(x) = p_1 x$ and a constant collision pressure with free parameters $p_1 < 0$ and $\cprs > 0$. We denote by $Y_B$ the base state solution of the yield-surface $y_B$.

\begin{figure}[ht!]
	\includegraphics[width=0.49\linewidth]{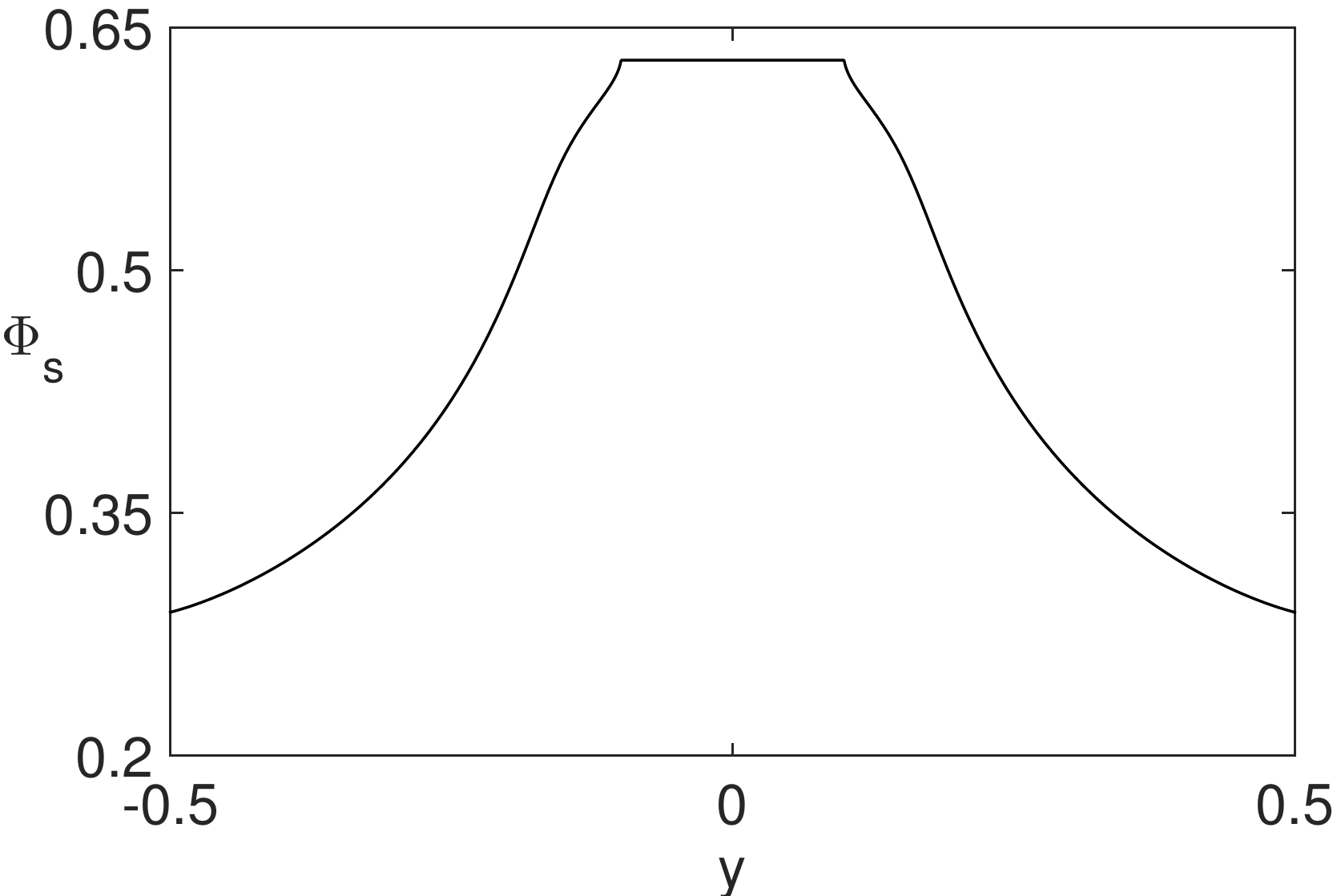}
	\includegraphics[width=0.47\linewidth]{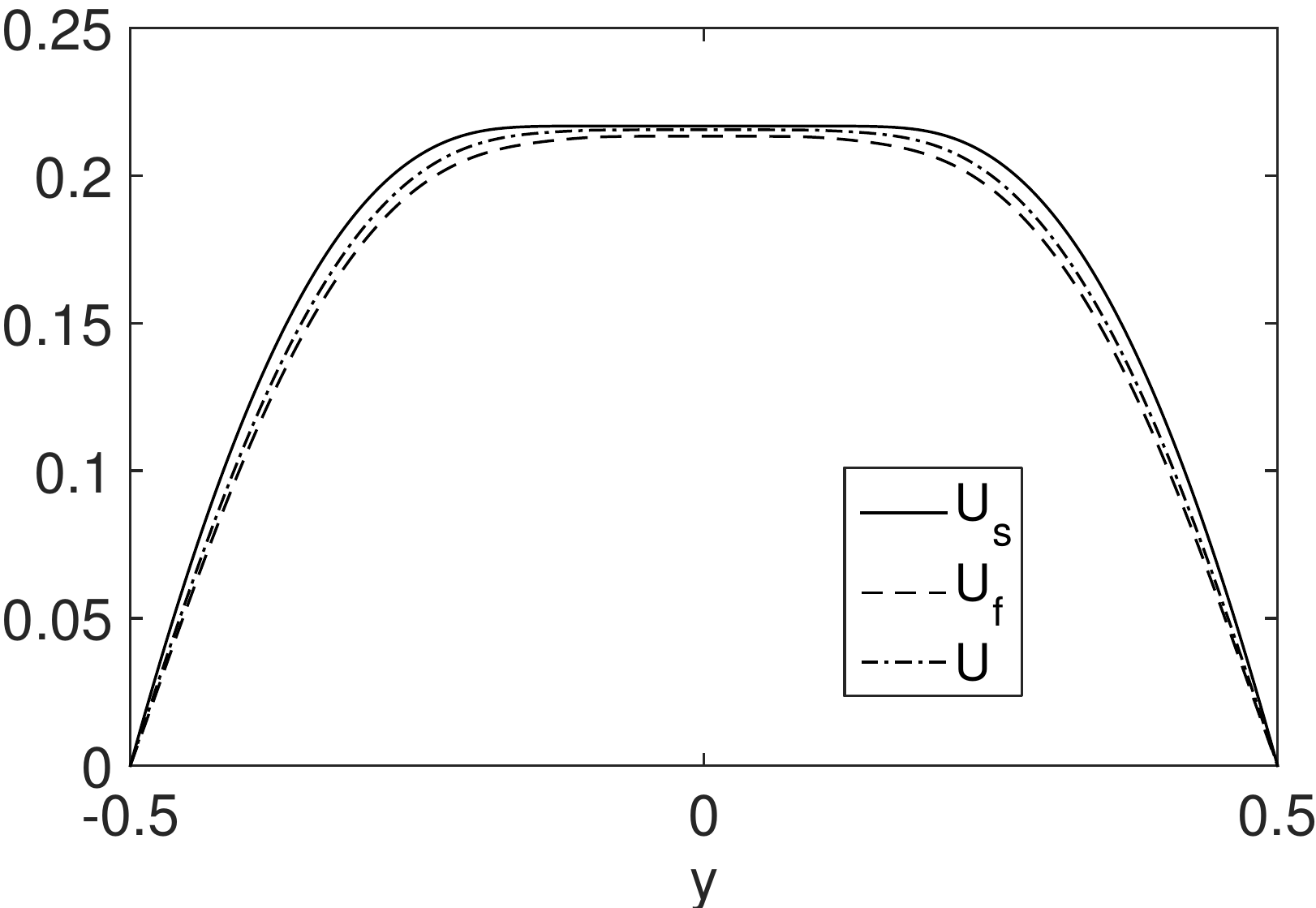}
	\caption{Shown is the multiphase base state with parameters chosen as $p_1 = -10$, $Da = 1000$, $I_0 = 0.005$, $\mu_1 = 1$,$\mu_2 = 1.5$, $\svfcrit = 0.63$ and $\cprs = 1$.}
	\label{fig:multiphase_base_state}
\end{figure}

In order to solve for the solid volume fraction and velocities, we use the transformation
\begin{align}
	y = \left(Y_B - \frac12\right) \zeta + \frac12,
\end{align}
define the shorthand notation
\begin{align}
	N(\svfb) \equiv \frac{\svfb\, \eta_s(\svfb)}{\eta_n(\svfb)},
\end{align}
and get the boundary value problem
\begin{subequations}
\begin{align}
	\frac{1}{Y_B - \frac12} \partial_{\zeta} \left(\frac{\left(\frac{1}{Y_B - \frac12} \partial_{\zeta} N + \svfb \, p_1 \right)(1-\svfb)}{\DA\,\svfb^2}\right) &= \frac{p_1 \left((Y_B - \frac12)\zeta + \frac12\right) + N}{1 - \svfb} + \frac{1}{\eta_n}, 
\end{align}
for the volume fraction base state $\svfb$ and $Y_B$ with boundary conditions
\begin{align}
	0 &= \partial_{\zeta} N + \left(Y_B - \frac12 \right) \svfb \, p_1 & \text{ at } \zeta &= 0, \\
	\svfb &= \svfcrit & \text{ at } \zeta &= 1, \label{eqn:dirichlet_transformed} \\
	\partial_{\zeta} \svfb &= -\frac{2(Y_B - \frac12)}{5(1 -\svfcrit)} \frac{\DA^{\frac12} \svfcrit (p_1 Y_B + \mu_1)}{\tanh \left(\frac{\DA^{\frac12} \svfcrit}{1 -\svfcrit} Y_B\right)} + \frac25 \left(Y_B - \frac12\right) p_1 & \text{ at } \zeta &= 1.
\end{align}
These results can be used in
\begin{align}
	p_c &= -\eta_n(\svfb) \partial_y \svelub, \\
	\fvelub &= \frac{(\partial_y N + \svfb p_1)(1 -\svfb)}{\DA\, \svfb^2} + \svelub,
\end{align}
for the fluid region $y > Y_B$ with no-slip boundary condition and
\begin{align}
	\svfb &= \svfcrit, \\
	\partial_y \svelub &= 0, \\
	\partial_y \fvelub &= \frac{p_1 y}{1 - \svfcrit},
\end{align}
in the plug-flow region with boundary conditions
\begin{align}
	\jumpop{\svelub} = \jumpop{\fvelub} &= 0 \qquad \text{ at } y = Y_B,
\end{align}
which yields the solution for the base states of the Poiseuille flow.
\end{subequations}
Figure \ref{fig:multiphase_base_state} shows an exemplary base state with a plug-flow region at the center of the channel. 

\subsubsection{Boundary conditions for the stability problem}\label{subsubsec:bcstab}

The linearized reduced two-phase system solves for the unknowns $\svfy, \svelvy, \sveluy$ and $\fvelvy$, where the last denotes the linearized $y$-component of velocity of the liquid phase for both - in the jammed and the liquid region. The corresponding equations have maximum orders of $0$, $2$, $2$, and $4 + 4$. Adding the free-boundary conditions at $y_B$, we get a minimum number of $13$ conditions.

The boundary condition for the plane Poiseuille flow are the no-slip boundary condition at the wall
\begin{align}
	\fvel = \svel &= \vc{0} \qquad \text{ at } y = 1/2, \label{eqn:poiseuille_no_slip}
\end{align}
symmetry around the center of the channel
\begin{align}
	\partial_y \fvel &= \vc{0} \qquad \text{ at } y = 0, \label{eqn:symmetry_condition}
\end{align}
and, following the example from the Bingham fluid, we impose continuity of the velocities and shear rates at the yield-surface
\begin{align}
	\llbracket \fvel \rrbracket = 
	\llbracket \svel \rrbracket =
	\llbracket \ssrt \cdot \vc{n} \rrbracket =
	\llbracket \fsrt \cdot \vc{n} \rrbracket &= \vc{0} \qquad \text{ at } y = y_B. \label{eqn:poiseuille_yield_surface_continuity}
\end{align}

Just as in the plane Couette flow case, cf.~\eqref{eqn:BCs_shear_case}, the no-slip conditions \eqref{eqn:poiseuille_no_slip} yield
\begin{align}
	\fvelvy = 
	\sveluy = 
	\svelvy &= 0 & \text{ and } &&
	\fvfb \partial_y \fvelvy + \svfb \partial_y \svelvy &= 0 \qquad \text{ at } y = 1/2. 
\end{align}

The symmetry condition \eqref{eqn:symmetry_condition} at the channel center yields
\begin{align}
	\partial_y \fvelvy &= 0 \quad \text{ at } y = 0.
\end{align}
Differentiation of equation \eqref{eqn:finished_system_conti_f} by $y$, the symmetry condition $\partial_y \fveluy = 0$ implies
\begin{align}
	\partial_{yy} \fvelvy &= 0 \quad \text{ at } y = 0.
\end{align}

For the conditions at the yield-surface $y = y_B$ we note that for any quantity $s$ with 
base state $S$ and  Fourier-transformed perturbation $\delta \hat s$,
linearizing a condition
\begin{align}
\llbracket s \rrbracket = 0 
\end{align}
at the yield surface leads to the expression 
\begin{align}
\llbracket \partial_y S \rrbracket \hp = -\llbracket \hat s \rrbracket,
\end{align}
where $y_b = Y_b + \delta \hp$.
Therefore, the continuity condition \eqref{eqn:poiseuille_yield_surface_continuity} gives
\begin{subequations}
\begin{align}
	\llbracket \tilde u_j \rrbracket = \llbracket \partial_y U_j \rrbracket \hp, \qquad
	\llbracket \tilde v_j \rrbracket = \llbracket \partial_y V_j \rrbracket \hp, \qquad
	\llbracket \tilde{\dot{\ts{\gamma}}}_s \rrbracket = \llbracket \partial_y \Gamma_s \rrbracket \hp, \qquad
	\llbracket \tilde{\dot{\ts{\gamma}}}_f \rrbracket = \llbracket \partial_y \Gamma_f \rrbracket \hp
\end{align}
and using the knowledge of the base states (e.g. continuity of $\partial_y \fvelub$), we obtain
\begin{align}
\llbracket \tilde u_j \rrbracket = 0, \qquad
\llbracket \tilde v_j \rrbracket  = 0, \qquad
\end{align}
\end{subequations}
at the yield surface $y = y_B$ for $j \in \{f,s\}$. 

This implies the boundary conditions
\begin{align}
	&& 
	\sveluy &= 0, &
	\svelvy &= 0, &
	\jumpop{\fvelvy} &= 0,
	&&
\end{align}
at the yield-surface $y = y_B$. We have, due to the continuum hypothesis \eqref{eqn:poiseuille_yield_surface_continuity} of the normal shear rates the representation
\begin{subequations}
\begin{align}
	\left\llbracket{ \begin{pmatrix}
		\partial_y \sveluy + i \alpha \svelvy \\
		\partial_y \svelvy
	\end{pmatrix} }\right\rrbracket &= -
	\left\llbracket{ \begin{pmatrix}
		\partial_{yy} \svelub \\
		0
	\end{pmatrix} }\right\rrbracket \hp, &
	\left\llbracket{ \begin{pmatrix}
		\partial_y \fveluy + i \alpha \fvelvy \\
		\partial_y \fvelvy
	\end{pmatrix} }\right\rrbracket &= -
	\left\llbracket{ \begin{pmatrix}
		\partial_{yy} \fvelub \\
		0
	\end{pmatrix} }\right\rrbracket \hp.
\end{align}
\end{subequations}
Due to $\ssrt = 0$ in $\Omega_s$, we have
\begin{subequations}
\begin{align}
	\partial_y \svelvy = 0, \qquad
	\jumpop{\partial_y \fvelvy} = 0 \quad \text{ at } y = y_B
\end{align}
as well as the free-boundary conditions
\begin{align}
	\jumpop{\partial_y \sveluy} &= -\jumpop{\partial_{yy} U_s} \hp \qquad \text{ at } y = y_B.
\end{align}
\end{subequations}
Using $\svelvy = \sveluy = \partial_y \svelvy = 0$ the solid transport equations yields
\begin{align}\label{qc333}
	\svfy &= 0 \quad \text{ at } y = y_B.
\end{align}

In summary, we have derived the required 13 conditions, i.e.~the wall boundary conditions
\begin{subequations}
\begin{align}
	&& \fvelvy = \sveluy = \svelvy &= 0, & \text{ and }&& 
	\fvfb \partial_y \fvelvy + \svfb \partial_y \svelvy &= 0, \quad \text{ at } y = 1/2, &&
\end{align}
the symmetry conditions
\begin{align}
	\partial_y \fvelvy &= \partial_{yy} \fvelvy = 0 \quad \text{ at } y = 0,
\end{align}
the yield-surface conditions
\begin{align}
	\sveluy &= \svelvy = 0,  \\
	\jumpop{\fvelvy} &= 0, \\
	\partial_y \svelvy &= \jumpop{\partial_y \fvelvy} = 0, \\
	\svfy &= 0,
\end{align}
at the plug-flow region boundary $y = y_B$ and the free-boundary condition
\begin{align}
	\llbracket \partial_y \sveluy \rrbracket &= -\llbracket \partial_{yy} U_s \rrbracket \hp, \qquad \text{ at } y = y_B.
\end{align}
\end{subequations}	

For the numerical investigations of the above model we combine our experience with the solution of the stability problem for the Couette flow problem as well as for the classic Bingham problem, and expand our finite-difference code to also deal with the singularity at the yield-surface in the two-phase Poiseuille flow. The employed scheme details are described in Appendix \ref{sec:NumericalScheme}.
We note first, that the two-phase Poiseuille flow also shows a collision pressure induced ill-posedness as well as a convection induced instability.

\subsubsection{Collision pressure induced ill-posedness}
\begin{figure}[t]
\includegraphics[width=0.99\linewidth]{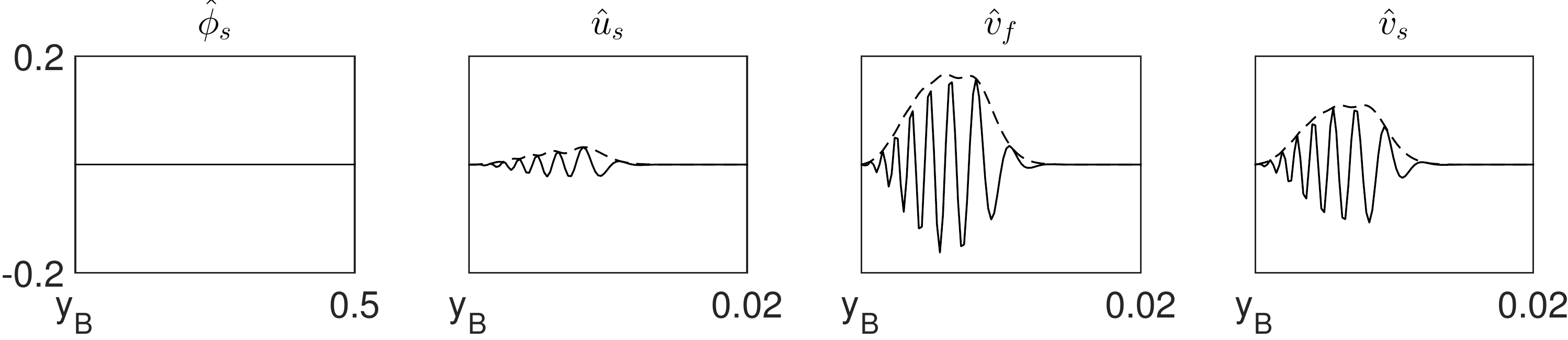}
\includegraphics[width=0.99\linewidth]{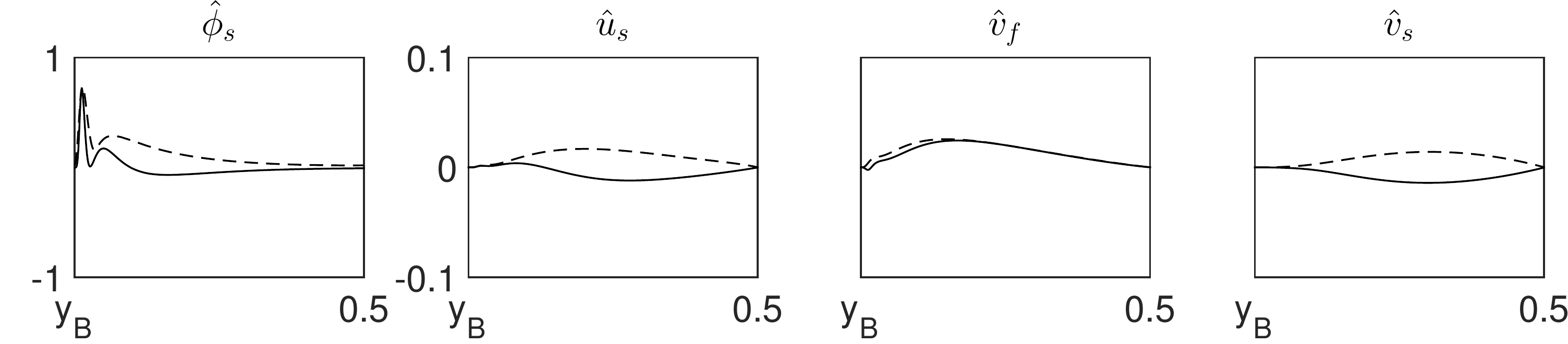}
	\caption{Top: A collision pressure induced growth mode for Poiseuille flow with parameter values $\DA=100, \RE=1, \phi_{sc}=0.63, p_c=1, p_1=-10, \alpha=1000, \mu_1=\mu_2=0.1$. Shown are the real value (solid line) and the absolute value (dashed line) of the mode. As for Couette flow the contribution of $\tilde\phi_s$ is negligibly small. The velocity modes spike  next to the yield surface $y_B$ and decays rapidly to zero towards the channel boundary $y=0.5$, shown here only until $y=0.02$. This demonstrates that the instability originates in the region of the highest particle concentration, as suggested by the analytic criterion \eqref{eqn:pc_instability_criterion}.
Bottom: A convection induced growth mode for Poiseuille flow with parameter values as above, except $\mu_1=\mu_2=1, \alpha=10$. In contrast to the collision pressure induced instability, $\tilde\phi_s$ exhibits the highest amplifications extending from the channel wall to the yield surface. To observe the small amplifications of the velocity modes we show only the region between [-0.1,0.1].}
\label{fig:instable_pc_mode_channel}
\end{figure}

The collision pressure induced ill-posedness from Section \ref{subsec:Collisionpressure} can be seen in numerical solutions starting at a ratio of $\svf \eta_s / \eta_n$ smaller than $1/4$. This is in contrast to the Couette flow, where the ill-posedness is already seen for a ratio of $1/2$ in the simulations. This can be explained by looking at the analytic criterion \eqref{eqn:pc_instability_criterion}, which shows that the ill-posedness occurs more likely in regions, where $\svf$ is close to maximum packing fraction. 
An unstable mode originates at the boundary of the plug-flow region, where the volume fraction is highest, but it is damped at the outer region, where the volume fraction is far from the maximum packing fraction. Figure \ref{fig:instable_pc_mode_channel} shows such a mode. Note the spike next to the plug-flow region, which shows that the growth is strongest there.
This suggests that the sufficient ratio between the viscosity of the solid phase and the collision pressure to suppress this ill-posedness depends on the base state. Thus, the normal mode analysis does not yield a sufficient criterion, as in \eqref{eqn:pc_instability_criterion} for general flows.

It is interesting to note that in two recent papers by Le Campion
et al.~\cite{Lecampion2014} and by Oh et al.\cite{Oh2015}, a similar
two-phase model for dense suspensions with a similar constitutive law
based on the work by Boyer et al.~\cite{Boyer2011} was constructed
and compared with experiments. In particular, it appears that the
authors were able to carry out numerical solutions without difficulties
arising from an ill-posedness in their model. Their simulations used values
for $\mu_1>1/4$ i.e.\ outside the range for which we found the
ill-posedness to occur. Moreover, their constitutive law allows for
compaction beyond the jamming limit, thus allow a variable density in the
unyielded region, which may further impact the stability properties.

\begin{figure}[ht!]
	\includegraphics[width=0.70\linewidth]{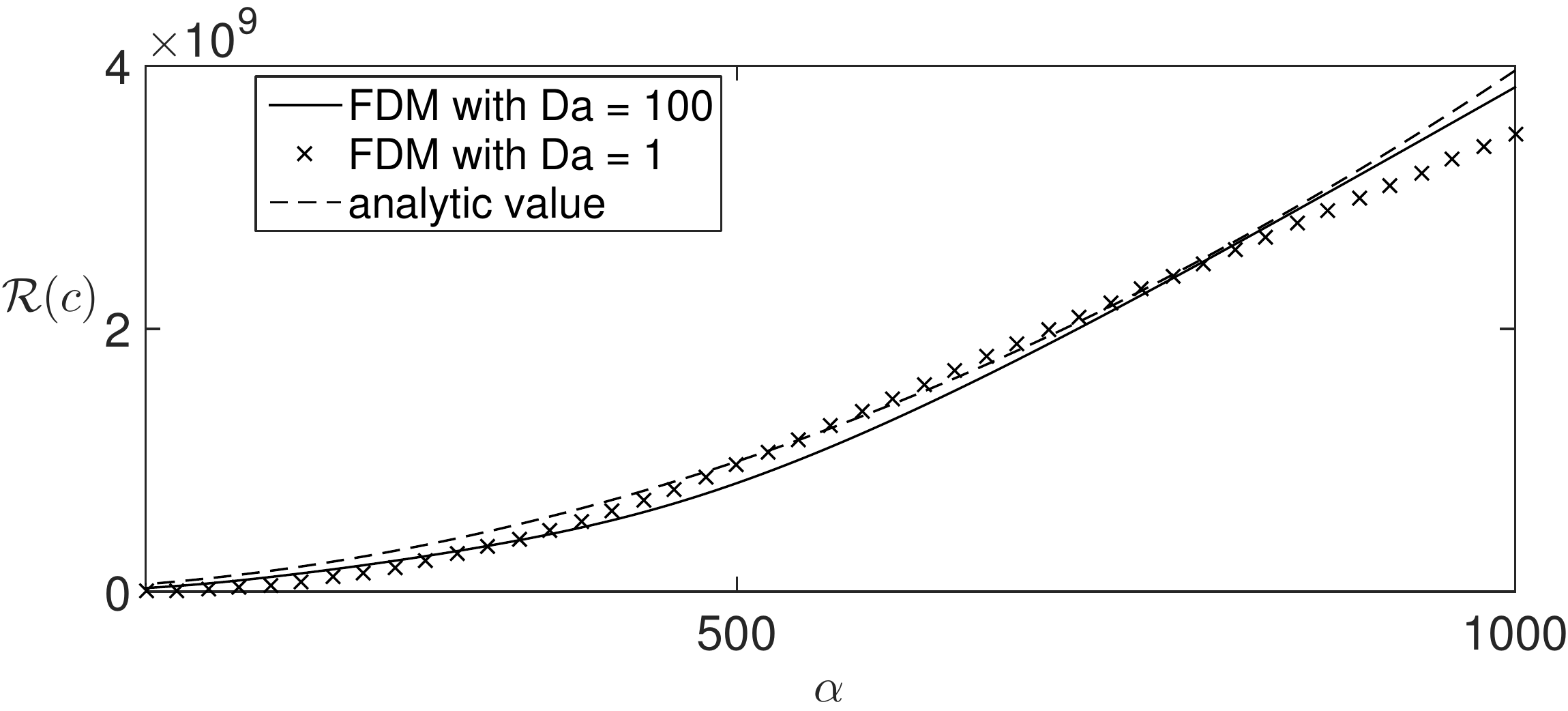}
		\caption{Shown is the dispersion relation of the collision pressure induced ill-posedness for the Poiseuille flow with parameters as in Figure \ref{fig:instable_pc_mode_channel}. The analytic curve is computed by equation \eqref{eqn:unstable_c_pc_full} with $\svfb = 0.62$, which has been only derived for the plane Couette flow. Since numeric and analytic results match well, we believe this instability has the same origin as explained in the plane Couette flow case.}
		\label{fig:dispersion_channel_collision}
\end{figure}
\subsubsection{Convection induced instability}
Unless $\mu_1$ is set too small, such that the collision pressure induced ill-posedness can be observed, unstable modes have real parts, which are of order one and have a similar signature as the convection induced unstable modes from Section \ref{subsec:Convection}. Figure \ref{fig:instable_pc_mode_channel} (Bottom) shows an exemplary unstable mode of that kind. Just as in the Couette flow case they appear in pairs and are strongest for the region between wall and plug-flow, where the velocities still change considerably, but $\svf$ is already near the maximum packing fraction. This is to be expected, since a high volume fraction and strong shearing are driving this instability.
We further note that large Reynolds and small Darcy numbers increase the convection induced instabilities, but seem not to introduce new instable modes for the Poiseuille flow case.
\subsection{Comparison of single- and multi-phase stability}
The single-phase Bingham flow and the multi-phase model showed different stability behavior. As discussed in Section \ref{subsec:Bingham}, the Bingham flow is unconditionally linearly stable when used with the correct boundary conditions. For the multi-phase model of Section \ref{subsec:TwoPhaseFlowModel} we found two instabilities: the collision pressure induced ill-posedness and the convection induced instability. 
However, the Bingham flow depends on only two parameters, i.e. the Reynolds number $\RE$ and the Bingham number $\mathrm{B}$. The Reynolds number arises in both models, but the Bingham number is just contained in the single phase model. As the Bingham number $\mathrm{B}$ has a direct influence on the size of the plug-region and the stress it plays a similar role as the viscosity of the solid phase $\eta_s$ and maximum packing parameter $\svfcrit$ in the multiphase mode. Yet, it seems to miss the ability to model the competition relative to the collision pressure $\eta_n$.

Both multi-phase model instabilities originate in mechanisms not contained in the single-phase model - the ill-posedness originates in the competition of the solid stress and solid pressure and the convection driven instability stems from the transport of particles due to convection. The Reynolds number does not play a significant role in either of the instabilities, which is similar to the single-phase model. 

\section{Conclusion}\label{sec:Conclusion}
In this work, we studied the stability properties of a multiphase model for concentrated suspensions for Couette and Poiseuille flow.
Our linear stability analysis showed two instabilities exhibited by the proposed model in case of plane Couette flow: a collision pressure driven ill-posedness and a convection induced instability. 
An analytic ansatz showed that the ill-posedness stems from a competition between the solid phase viscosity and the collision pressure and poses a necessary stability condition on the size of the solid phase viscosity compared to the collision pressure. This has been reaffirmed by comparison between numerical and analytical results. 

The convection driven instability has been analyzed using a Kelvin-mode ansatz. The resulting time dependent ordinary differential equations showed a transient instability.  We note that this might prohibit an experiment from showing the Couette or Poiseuille flow base state, because of the onset of turbulence or the occurrence of shocks for highly concentrated suspensions.
The consequence of the convection driven instability for the studied base states can best be analyzed using a direct numerical simulation of the full model, which will be part of our future work. 
In addition, further extensions of the underlying model that include intermolecular short range forces may become necessary for the stability properties that relate to the formation of anisotropic microstructures for high enough volume fractions, see e.g. \cite{vazquez2016rheology,gadala1980shear}.

In case of the Poiseuille flow, we also retrieved the multi-phase instabilities and compared the multiphase model to the stability of the Bingham flow. 
We also note here, that since Poiseuille flow for our two-phase model contains  different velocities of the solid and liquid phase, the problem of loss-of-hyperbolicity might arise here too. 
Our numerical studies therefore focused on cases with large velocity differences between solid and liquid phases, as one would expect this transition to occur in those cases. However, as has been shown in \cite{Prosperetti1987, Stewart1979}, while the loss-of-hyperbolicity and the associated ill-posedness can only be observed in the long-wave limit and additionally with fine meshes,  our numerical results did not yield new unstable modes, even for rather small wave numbers, such as $\alpha < 0.01$. It remains to be shown if this picture changes for higher resolutions, smaller viscosity terms or perhaps also different base states. It would thus be interesting to see if for our two-phase flow model the ill-posedness can also be connected to the existence of a singular shock, such as has been seen in applications detailed in Carpio et al.~\cite{Carpio2001} or Bell etal.~\cite{Bell1986}, or in connection with other operators studied by Zhou et al. ~\cite{Zhou2005} and Cook et al.~\cite{Cook2008}. 

\section*{Acknowledgements}

TA gratefully acknowledges the support by the Federal Ministry of Education (BMBF) and the state government of Berlin (SENBWF) in the framework of the program Spitzenforschung und Innovation in den Neuen L\"andern (Grant Number 03IS2151).

\begin{appendix}

\section{Bingham-Orr-Sommerfeld boundary conditions}\label{sec:BinghamBCs}

The Bingham-Orr-Sommerfeld equation is obtained by integrating the linearized problem about the base state 
\begin{align}
	U_B(y) &=
	\begin{cases}
		 1 - \frac{(|y|- y_B)^2}{(1/2 - y_B)^2}  & \text{ for } y_B \le |y| \le 1/2 \\
		 1 & \text{ for } |y| < y_B
	\end{cases}
\end{align}
which results from making the ansatz 
$\vel = (U_B(y),V_B(y))$, $p = P x$,
and split the domain into a plug-flow and a fluid region, i.e. $\Omega = \Omega_f \cup \Omega_s$.

For the linearization we set $\vel = \vc{U}_B + \delta \velp$ and define 
\begin{align}
	\eta(\vc{u}) = \frac{1}{\mathrm{Re}}\left( 1 + \frac{\mathrm{B}}{|\srt|}\right).\nn
\end{align}
Then
\begin{align}
	\eta(\vc{U}_B + \delta \velp) &=  \eta(\vc{U}_B) 
	- \delta  \frac{1}{2} \sum_{i,j} \srtcmpt_{ij}(\velp)\srtcmpt_{ij}(\vc{U}_B) \frac{\mathrm{B}}{\mathrm{Re} \; |\srt(\vc{U}_B)|^3}
	+ O(\delta^2)\nonumber,\\
	\stresscmpt_{ij}(\vc{U}_B + \delta \velp) &= \stresscmpt_{ij}(\velb) 
	+ \delta \eta' \srtcmpt_{ij}(\velb) + \eta(\velb) \srtcmpt_{ij}(\velp)
	+ O(\delta^2)\nonumber,\\
	|\stress(\velb + \delta \velp)| &= |\stress(\velb)| + \delta \frac{1}{2} \frac{\sum_{i,j} \stresscmpt_{ij}'(\velp) \stresscmpt_{ij}(\velb)}{|\stress(\velb)|} + O(\delta^2).\nn
\end{align}
together with the perturbed yield criterion $H = \pm y_b \pm \delta h$ for the position of the yield surface results to $O(\delta)$ in a linerized sytem, that can be integrated using the ansatz 
$(\velup,\velvp,\prsp) = (\hat u(y), \hat v(y), \hat p(y)) \mathrm{e}^{i\alpha(x - ct)}$ to obtain 
gives the Orr-Sommerfeld-Bingham equation
\begin{align}
	i \alpha \mathrm{Re} [(U_B - c) (\partial_{yy}\hat{v} - \alpha^2 \hat{v}) - \hat{v}\,\partial_{yy} U_{B}] &= \left( \partial_{yy} - \alpha^2 \right)^2 \hat{v} - 4 \alpha^2 \mathrm{B}\, \partial_y \left( \frac{\partial_y\hat{v}}{|\partial_y U_{B}|} \right).
	\label{eqn:final_orr_sommerfeld_bingham}
\end{align}

For the boundary conditions we have the no-slip boundary condition at the wall 
\begin{align}
	\velvy(\pm 1/2) &= 0, & 
	\partial_y \velvy(\pm 1/2) &= 0.\nn
\end{align}
At the yield surface, we have due to the normal shear rate continuity
\begin{align}
	\partial_y\velvp(x,\pm y_B, t) &= 0, \nn\\
	\partial_y\velup(x,\pm y_B, t) + \partial_x\velvp(x,\pm y_B, t) &= \mp\, h \,\partial_y \srtcmpt_{12}(\velb, \pm y_b) = \frac{\pm 2 h}{(1/2 - y_B)^2},\nn
\end{align}
which yields with normal mode ansatz and usage of the continuum equation
\begin{align}
	\partial_y \velvy &= 0, &
	\partial_{yy} \velvy - \alpha^2 \velvy &= \frac{\mp i\alpha 2h}{(1/2 - y_B)^2}.\nn
\end{align}

In the plug-flow bulk region $(x,y) \in \Omega_s$, we have to order $O(\delta)$ and using the normal modes ansatz 
\begin{align}
	\veluy(x,y) = 0, \quad
	\partial_y\,\velvy(x,y) = 0, \quad
	\velvy(x,y) = 0.\nn
\end{align}
Continuity of $\vel$ at the yield surface and using $\velb^+(y_B) = \velb^-(y_B)$, we get
\begin{align}
	 \hp\,\partial_y\left( \velb^+(x,y_B) - \velb^-(x,y_B)\right) +  (\velp^+(x,y_B) - \velp^-(x,y_B)) = 0.\nn
\end{align}
We have $\partial_y\left(\velb^+(x,y_B) - \velb^-(x,y_B)\right) = 0$, so
\begin{align}
	\velp^+(x,y_B) = \velp^-(x,y_B),\nn
\end{align}
and since $\velp^- =  \vely^-(y) \mathrm{e}^{i \alpha (x - ct)} = 0$ due to $\vely^-(y) = 0$, we have
\begin{align}
	\velp^+(x,y_B) = 0.\nn
\end{align}

Overall we have the boundary conditions
\begin{subequations}
	\label{eqn:single_phase_stability_BCs}
	\begin{align}
		\hat{v} &= \partial_y\hat{v} = 0 & \text{ at } & y = 1/2, \\
		\hat{v} &= \partial_y\hat{v}= 0 & \text{ at } & y = y_B, \\
		\partial_{yy}\hat{v} &= \frac{-i\alpha 2 h}{(1/2 - y_B)^2} & \text{ at } & y = y_B.
	\end{align}
\end{subequations}

\section{Numerical scheme}\label{sec:NumericalScheme}

The basic idea of the scheme is to use neighboring half points for the approximation of the derivatives. Suppose we have mesh points $x_i \in \mathbb{R}$ with constant width $h = x_{i+1} - x_i$ and suppose we have a function $f(x)$ with $f_i := f(x_i)$. Let us further define the half-points $x_{i+1/2} := (x_i + x_{i+1})/2$ and $f_{i+1/2} := (f_{i+1} + f_i)/2$ then we define the discrete derivatives as 
\begin{subequations}
\begin{align}
	\nabla_h f_{i+1/2} &= (f_{i+1} - f_{i})/h, \\
	\nabla_h f_i &= (f_{i+1/2} - f_{i-1/2})/h &=& (f_{i+1} - f_{i-1})/(2 h), \\
	\nabla^2_h f_i &= (\nabla_h f_{i+1/2} - \nabla_h f_{i-1/2})/h &=& (f_{i+1} - 2f_i + f_{i-1})/h^2, \\
	\nabla^2_h f_{i+1/2} &= (\nabla_h f_{i+1} - \nabla_h f_{i})/h &=& (f_{i+2} - f_{i+1} - f_i + f_{i-1})/(2 h^2), \\
	\nabla^3_h f_{i} &= (\nabla^2_h f_{i+1/2} - \nabla^2_h f_{i-1/2})/h &=& (1/2 f_{i+2} - f_{i+1} - f_{i-1} + 1/2 f_{i-2})/h^3, \\
	\nabla^3_h f_{i+1/2} &= (\nabla^2_h f_{i+1} - \nabla^2_h f_{i})/h &=& (f_{i+2} - 3 f_{i+1} + 3f_{i} - f_{i-1})/h^3, \\
	\nabla^4_h f_{i} &= (\nabla^3_h f_{i+1/2} - \nabla^3_h f_{i-1/2})/h &=& (f_{i+2} - 4 f_{i+1} +6 f_i - 4f_{i-1} + f_{i-2})/h^4,
\end{align}
which is just the standard central scheme of second order for entire points.
\end{subequations}

For the multiphase model we additionally used a staggered grid scheme, where the velocities $\sveluy,\fvelvy,\svelvy$ live on entire points and the volume fraction on half points, i.e. $\sveluy{}_i := \sveluy(x_i)$ and $\svfy{}_j := \svfy(x_{j+1/2})$.
This approach evades a decoupling of odd and even points in the volume fraction, that has been observed when using the standard central scheme for the transport equation \eqref{eqn:final_multiphase_conti_s} in the multiphase model.

After discretization of system \eqref{eqn:final_fluid_reduced} and possibly equation \eqref{eqn:final_solid_reduced}, we receive two matrices. The first matrix contains the spatial derivatives and the second matrix the discretization for the time mode $c$, so that we get a system of the form
\begin{align}
	\ts{A} \vc{v} = c \ts{B} \vc{v},
\end{align}
which has been solved using the generalized eigenvalue solvers in Matlab.

The boundary conditions are implemented using the ghost-point method and they are explicitly eliminated before solving the generalized eigenvalue problems. This circumvents the appearance of pseudo-eigenvalues stemming from the ghost-points, which can be of any value, even infinity and do not give new insight into the stability of the system.

As the system is complex and its implementation prone to errors, we looked for a possible validation method. 
We first tested our scheme for the Newtonian Couette-flow problem leading to the corresponding well-studied Orr-Sommerfeld equation~\cite{Orszag1971} as well as for the non-Newtonian case leading to the Orr-Sommerfeld-Bingham equation~\cite{Frigaard1994}.

For another independent validation we neglect the convective term $\partial_x(\svelub \svfy)$ and set $\RE = 0$. Then the Couette flow permits an analytic solution. Using the Fourier ansatz $e^{ikx + i\ell y - i m t}$ in system \eqref{eqn:multiphase_linearized}, we are able to derive an algebraic system. The derived algebraic system and the numerical approximation show excellent agreement.

\end{appendix}

\bibliography{AMNW-StabSuspension}
\bibliographystyle{unsrtnat}

\end{document}